\documentclass[a4paper,11pt]{article}
\usepackage{pos}

\def\barray{\begin{array}}
\def\earray{\end{array}}
\def\be{\begin{equation}}
\def\ee{\end{equation}}
\def\bea{\begin{eqnarray}}
\def\eea{\end{eqnarray}}
\def\bal{\begin{align}}
\def\eal{\end{align}}

\def\({\left(}
\def\){\right)}
\def\[{\left[}
\def\]{\right]}

\title{Local supersymmetries and three-charge black holes}

\author*[a]{Yixuan Li}

\affiliation[a]{Max-Planck-Institut f\"ur Physik (Werner-Heisenberg-Institut), \\
	F\"ohringer Ring 6, 80805 M\"unchen, Germany}

\emailAdd{yixuan@mpp.mpg.de}

\abstract{
Local supersymmetry enhancement (LSE) is a formalism intended to identify information needed to describe microstates of supersymmetric black holes that are realised as brane systems. 
After illustrating the relationship between LSE and black-hole microstates with the F1-P system, we review two possible strategies to apply the LSE mechanism to get microstates of three-charge black holes. While the former leads to microstates that break the spherical symmetry of the horizon, microstates built from the latter preserve it, and are numerous enough to account for (at least) a finite fraction of the M2-M5-P black-hole entropy. This proceeding is partially based on \cite{Bena:2022wpl} and on unpublished notes. 
}

\FullConference{%
  Corfu Summer Institute 2022 "School and Workshops on Elementary Particle Physics and Gravity",\\
  28 August - 1 October, 2022\\
  Corfu, Greece}

\begin{document}
\tableofcontents
\maketitle

\section{Introduction: horizons and branes}

One of the greatest paradigm shifts introduced by string theory is that one can understand black holes with the dynamics of the branes. The key point is that branes source both gravitational and gauge fields. When one combines different types of branes in different configurations, the gravitational field can be superposed, and the resulting geometry is described by the metric, in the asymptotic observer's coordinates, of a black hole.

A very familiar example is the D1-D5-P system. In this system, the D1 and D5 branes wrap a common compact direction, $S^1$, and the D5 brane wraps in addition a four-dimensional compact manifold, for instance $T^4$ or $K3$. In addition, this system contains gravitational waves propagating in one of the directions of the circle $S^1$. See Table \ref{tab:D1-D5-P}.
\begin{table}[h]
    \centering
    \begin{tabular}{|c||c|c|c|c|}
    \hline
       & $t$ & $\mathbb{R}^4$ & $S^1_y$       & $T^4$    \\ \hline \hline
    D5 & $-$ & $\bullet$        & $-$            & $-$     \\ \hline
    D1 & $-$ & $\bullet$        & $-$            & $\sim$ \\ \hline
    P  & $-$ & $\bullet$        & $\rightarrow$ & $\sim$ \\ \hline
    \end{tabular}
    \caption{Brane configuration of the D1-D5-P system. Here, we use the convention where $-$ indicates that the brane/string is extended in the given dimension(s), $\bullet$ indicates that it is pointlike, and $\sim$ indicates that the brane is smeared in the given dimension(s). The arrow $\rightarrow$ indicates that the gravitational wave P is moving in one of the directions (left or right) of the circle, $S^1_y$. }
    \label{tab:D1-D5-P}
\end{table}

The supergravity solution coming from the brane system at Table \ref{tab:D1-D5-P} is determined through the \textit{harmonic-function rules}:
\begin{align} \label{eq:D1-D5-P_metric}
ds^2 &=  \frac{1}{\sqrt{H_1H_5}}  \left[ -dt^2+ dy^2 + \left( H_P -1 \right) (dt+dy)^2 \right] 
+ \sqrt{H_1H_5} \, ds^2_{\mathbb{R}^4} + (H_1H_5)^{-1/2}ds^2_{T^4}\,,\\
e^{2\phi} &= \frac{H_1}{H_5} \,, \label{eq:D1-D5-P_dilaton}
\end{align}
where 
\begin{align} \label{harmonic_functions}
    H_{1,5,P}=1+\frac{Q_{1,5,P}}{r^2} \,, \qquad r \equiv r_{\mathbb{R}^4} \,.
\end{align}
The supergravity charges can be expressed in terms of the number of branes and momentum quanta:
\be \label{eq:sugra_charges_vs_number_branes}
Q_1= \frac{g_s \alpha'}{v} N_1 \,, \qquad
Q_5= g_s \alpha' \, N_5 \,, \qquad
Q_P= \frac{(g_s)^2 \alpha'}{v \rho_y^2} N_P \,,
\ee
where $v\equiv \frac{V_4}{(2\pi)^4\alpha'^2}$ is the volume of the four-torus $T^4$ measured in units of $2\pi l_s$ and $\rho_y\equiv \frac{R_y}{l_s}$ is the radius of the $y$ circle measured in units of $l_s$.

It is important to stress here that harmonic-function rule that determined the metric \eqref{eq:D1-D5-P_metric} uses the fact that the D1 branes and the momentum quanta are \textit{smeared} on the $T^4$: The D1 branes (and the momenta quanta) are in principle points in the $T^4$ directions, but when they form a regular and densely packed array in $T^4$, one can \textit{approximate} the array by an uniform distribution of D1 branes (and momenta quanta) along the $T^4$.%
\footnote{Without smearing, D1 branes in a $\mathbb{R}^{8,1}\times S^1$ topology sources the string-frame metric 
\be
ds^2= H_1(r)^{-1/2}(dt^2+dy^2) + H_1(r)^{1/2}ds_{\mathbb{R}^8}^2 \,, \qquad 
H_1(r) = 1+ \frac{Q'_1}{\hat{r}^6} \,,
\ee
where $\hat{r}$ measures the distance to the D1-brane source in $\mathbb{R}^{8}$. One can apply the harmonic-function rule to get \eqref{eq:D1-D5-P_metric} only with the smeared harmonic functions \eqref{harmonic_functions} which all behave as $\frac{1}{r^2}$ and only depend on the overall transverse coordinates, $\mathbb{R}^4$.
}
For more detail, see for example \cite{Peet:2000hn} and references therein. The point is that when the D1 branes are smeared along the $T^4$ directions, one does not know where exactly they are localised in the $T^4$.

The solution (\ref{eq:D1-D5-P_metric}), (\ref{eq:D1-D5-P_dilaton}) can be reduced to a 5-dimensional metric in the Einstein frame:
\be
ds^2=\left(H_1H_5H_P\right)^{-2/3}dt^2 
+ \left(H_1H_5H_P\right)^{1/3}\left[ dr^2+r^2 d\Omega_3^2 \right] \,.
\ee
This is the metric of a 5-dimensional black hole. The horizon, determined by the equation $g^{rr}=0$, lies at $r=0$.
The area of the horizon, measured in the five-dimensional Einstein frame, is determined by the number of branes and momenta quanta:
\be \label{area_5_Q1Q5QP}
A_{(5)}=\int d\Omega_3 \, r^3 \( H_1 H_5 H_P \)^{3/6} \propto \sqrt{Q_1Q_5Q_P}
\propto g_s^2 (l_s)^3 \sqrt{N_1N_5N_P} \,.
\ee
The near-horizon geometry has an $\mathrm{AdS}_2 \times S^3$ geometry.

Actually, a crucial step in the equation \eqref{area_5_Q1Q5QP} is that the harmonic functions have the form $H_i \sim \frac{Q_i}{r^2}$ in the region $r\ll Q_i$. If the harmonic functions had a different behaviour close to $r=0$, say $H_i \sim \frac{1}{r}$, one would get $A_{(5)}=0$.


\subsection{Local supersymmetry enhancement: a first example}

\label{ssec:LSE_first_ex}

The lesson we learn from the Introduction is that for brane/strings/momenta that are compatible for their supersymmetry, one can apply the harmonic-function rule. For a 1/8-BPS system (like the D1-D5-P system), this rule gives a black-hole solution in supergravity, with a macroscopic horizon, whose area is determined by the asymptotic supergravity charges (\ref{eq:sugra_charges_vs_number_branes}).

A possible conclusion that could be drawn from these lessons is that: \vspace{0.3em}\\
\textit{The asymptotic charges and supersymmetries of the branes/stings/momenta seem to control the near-horizon geometry, all the way to the horizon. Therefore, whatever the microscopics of the brane system that could be distinguished at weak coupling ($g_sN\ll 1$), one always ends up with the same horizon in the supergravity ($g_sN\gg 1$, $g_s\ll1$) regime.
To have access the information about the microstates however, one would need to un-smear the brane system at $g_sN\ll 1$ and localise the momentum carriers; in the regime of gravity, because all microstates behave the same at the horizon region, it could only mean to probe the region near the singularity (string scale away from the singularity) -- and this region is precisely where supergravity breaks down. Therefore, it seems impossible that supergravity is able to probe the physics of black-hole microstates.}

\vspace{1em}
But are we sure that if one un-smears the momentum carriers in the open-string regime ($g_sN\ll 1$), one gets the same horizon in the supergravity regime? The reasoning rests on the standard lore saying that the only way to get a supergravity solution is to use the harmonic-function rule, which relies on the compatibility of supersymmetries. However, there is a more general notion that enable the use of the harmonic-function rule: \textit{local supersymmetries}. 

The archetypical example of local supersymmetries is that of the F1-P system. 
Consider the combination of a fundamental string F1 wrapped along a compact direction, $y$, and a momentum P along the same direction.
Alone, the string F1$(y)$ preserves 16 real supercharges, and so does the momentum P$(y)$. But the supercharges they preserve are not all the same, and 8 of the 16 supercharges are still preserved by F1$(y)$ and P$(y)$. One can thus apply the harmonic-function rule to the F1-P system and get a metric that depends on the overall orthogonal directions, $\mathbb{R}^8$.

\begin{figure}[h]
\centering
\includegraphics[width=0.8\linewidth]{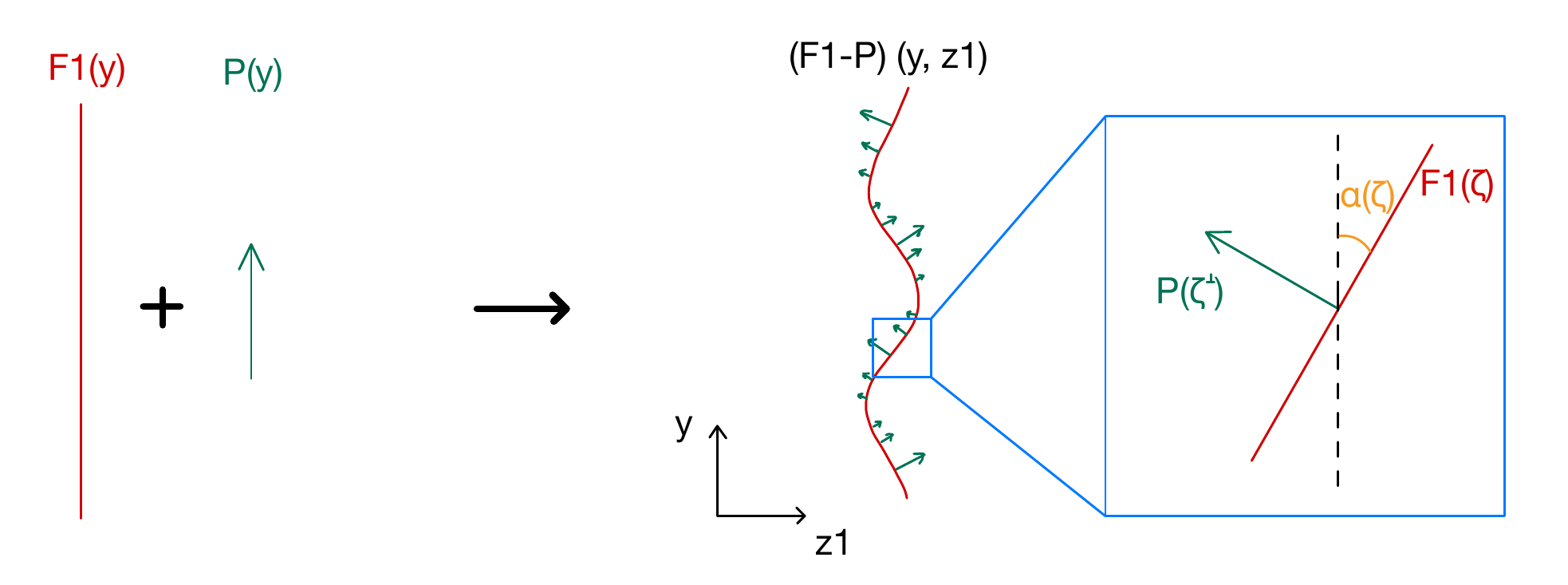}
\caption{Schematic depiction of the combination of F1$(y)$ with P$(y)$ (left) and that of a string carrying momentum through its transverse oscillations (right). The string worldline, along $\zeta$, makes an angle $\alpha(\zeta)$ with the direction $y$.}
\label{fig:F1-P_wiggle}
\end{figure}

However from the microscopic point of view, the only way for a string to carry momentum is to have transverse oscillations. The classical picture of this microscopic realisation is that the string carries longitudinal momentum by having a profile in its orthogonal directions, in $\mathbb{R}^8$. To be more precise, the oscillations move at the speed of light along $y$ and thus carry momentum, as the profile is independent from one of the two null-like coordinates on the circle $S^1_y$, that we call $u$. The profile varies along the string's worldline, $\zeta$, and can be parameterised by an angle, $\alpha(\zeta)$. See Fig. \ref{fig:F1-P_wiggle}. 
The F1-P profile preserves the same eight global supersymmetries as those preserved by the F1$(y)$ and P$(y)$. But locally, the F1-P profile is a piece of a fundamental string along a direction $\hat{y}$, boosted along an orthogonal direction, $\hat{y}^\perp$; this local boosted fundamental string preserves 16 supersymmetries.


In a nutshell, given the F1$(y)$ and P$(y)$ charges which preserve 8 supersymmetries together, there is a way to realise this system microscopically as a string carrying momentum by having orthogonal oscillations, so that the number of local supersymmetries is enhanced to 16. To have 16 local supersymmetries is a sign that the degrees of freedom involved in the microscopic resolution of the F1-P system by the transverse oscillations are fundamental degrees of freedom in string theory. And the crux of the matter is that one can apply the harmonic-function rule for a local piece of this boosted fundamental string which has 16 supersymmetries and get a supergravity solution for a given global profile.

\vspace{1em}
The remaining of the proceeding organises as follows. In Section \ref{sec:global_local_susies}, we first develop the general formalism for local supersymmetries. We then illustrate it with the example of the F1-P system, and discuss its microscopic realisation as microstate geometries. We then review the logic behind the construction of microstate geometries for three-charge black holes.
In Section \ref{sec:internal_excitations}, we find the local supersymmetries for the  F1-NS5-P or M2-M5-P system, and explain what it implies at the level of the microstates. We conclude in Section \ref{sec:conclusion}. Finally, in Section \ref{sec:questions}, we identify a list of future directions in a ``Q \& A'' session.

\section{Global and local supersymmetries}
\label{sec:global_local_susies}

\subsection{The general formalism}

Type II String Theory vacua preserve 32 supersymmetries. Adding \textit{excitations} such as strings, branes or momentum waves decreases the number of preserved supersymmetries. Indeed, one can derive, using the BPS equations, that the presence of branes imposes a constraint on the Killing spinor $\epsilon$:
\begin{equation}
 	P \, \epsilon ~=~ -\epsilon \,,\qquad \text{or equivalently} \qquad \Pi \, \epsilon ~\equiv~ \frac12 (1 + P) \, \epsilon ~=~ 0\,,
 	\label{eq:bps_constraint_simple}
 \end{equation} 
 where $P$ is a traceless {\em involution} ($P^2 = 1$), typically a product of gamma matrices and Pauli matrices, that depend on the exact type and orientation of the object considered. Equivalently, $\Pi$ defined in \eqref{eq:bps_constraint_simple} is a projector, verifying $\Pi^2 = \Pi$. A list of the involutions corresponding to branes, strings, solitons and momentum waves is given in Appendix \ref{sec:projectors_and_involutions_for_branes}. 
 The constraint \eqref{eq:bps_constraint_simple} divides the number of preserved global supersymmetries by two.%
\footnote{$P^2=1$, and $\mathrm{tr}(P)=0$, so half of $P$'s eigenvalues are $+1$, half of them are $-1$.}

If one considers configurations with several types of branes whose supersymmetries are compatible, the constraints add up. For example, for a two-charge system, the Killing spinor must respect
\begin{equation}
	\Pi_1 \, \epsilon ~=~ 0 \,,\quad \text{and} \quad \Pi_2 \, \epsilon ~=~ 0 \,.
\end{equation}
Thus, the Killing spinor must lie in the intersection of the kernels of $\Pi_1$ and $\Pi_2$. And the dimension of this intersection is the number of preserved global supersymmetries (for the F1-P system, it is 8).

More generally, let's combine $k$ different excitations. Then
\be 
\epsilon \in \ker(\Pi_1) \cap \ldots \cap \ker(\Pi_k) \,.
\ee 
Now, let us consider other excitations, corresponding to involutions $(P_{k+1},\dots,P_n)$, and a set of weights $\alpha_1,\dots,\alpha_n >0$ such that $\alpha_1+\ldots+\alpha_n=1$.
Consider the matrix
\begin{equation}
	\hat\Pi  ~\equiv~ \frac12 \left(1 + \alpha_1 P_1 + \dots + \alpha_n P_n \right) ~=~ 0 \,.
\end{equation}
For each species of excitation, $i$, the coefficient $\alpha_i$ is the ratio between the charge density corresponding to this brane, $Q_i$,\footnote{Note that the dependence in the string coupling, $g_s$, enters in the $Q_i$'s.} and the mass density of the full bound state, $M$:
\begin{equation}
\displaystyle
\alpha_i \equiv \frac{Q_i}{ M}  \,. \label{QMratio}
\end{equation}
Hence, the projector can be written as 
\begin{equation}
	 \hat\Pi  ~=~ \frac{1}{2 M} \(M + Q_1 P_1 + \ldots + Q_n P_n\) \,.
\end{equation}
The matrix $\hat\Pi$ represents a mix of a set of $n$ different types of excitations. The overall mass of the mix is $M$ and each type of excitation contributes to a charge $Q_i$.

$\hat{\Pi}$ is not necessarily a projector. But $\hat{\Pi}$ is a projector if and only if the mix of excitations defined by it is a bound state (a state with 16 supersymmetries). Note that one does not necessarily need to enhance to 16 supersymmetries: One can, for example, enhance the supersymmetries from 4 to 8.

So, the procedure to enhance the supersymmetries of the original excitations ($i\in \{1,\dots,k\}$) is to: (1) consider in addition some auxiliary excitations ($i\in \{k+1,\dots,n\}$) and then (2) find the charge-to-mass ratios $\alpha_i$ so that $\hat\Pi$'s eigenvalue 1 gets more degeneracy. This is supersymmetry enhancement. But what is \textit{local} supersymmetry enhancement?

Given the involutions $P_i$ ($i\in \{1,\dots,n\}$), the set of solutions $\{\alpha_i\}$ is not unique. Therefore, they can depend on spacetime coordinates, $x$: $\{\alpha_i(x)\}$. Since this formalism -- at the level of projectors corresponding to the branes, strings, etc. -- corresponds to reading off the leading order of the Killing spinor equations in orthogonal coordinates to the bound state, $x$ here can only label coordinates on the brane bound state.%
\footnote{For the F1-P bound state for example, the coordinate $x$ labels the `spatial' part of the wiggling string's worldline. For a NS5-F1 bound state, $x$ labels a five-dimensional spatial worldvolume coordinates.}

Now, we are looking for solutions $\epsilon(x)$ to the equation 
\be 
\hat\Pi(x) \, \epsilon(x)  = 0 \,, \quad \forall x\,. 
\label{eq:susy_constraint_all_x}
\ee
At a given $x$, $\epsilon(x)$ is an element of $\ker(\hat \Pi(x))$, whose dimension gives the number of \textit{local supersymmetries}.

But globally, a Killing spinor $\epsilon_0$ preserved by all local pieces of the bound state should satisfy
\be \epsilon \in \bigcap_{x} \ker\left(\hat\Pi(x)\right) \,. \ee
The last vector space is required to match $\ker(\Pi_1) \cap \ldots \cap \ker(\Pi_k)$, so that one finds the original global supersymmetries one started with. This is ensured by writing the projector $\hat\Pi$ in the form
 \begin{equation}
 	\hat\Pi(x) ~=~ f_1(x) \, \Pi_1 + \ldots + f_k(x) \, \Pi_k \,,
 	\label{eq:global_susy_general}
 \end{equation}
where $\Pi_1, \dots, \Pi_k$ are the commuting projectors we started with, and $f_1, \dots, f_k$ can be any matrix-valued functions. Then, satisfying \eqref{eq:susy_constraint_all_x} is equivalent to
 \begin{equation}
 	\Pi_1 \, \epsilon ~=~ \ldots ~=~ \Pi_k \, \epsilon ~=~ 0\,,
 \end{equation}

At a given point $x_1$, $\ker\left(\hat\Pi(x_1)\right) \supset \bigcap_{x} \ker\left(\hat\Pi(x)\right)$, so we immediately deduce that the number of local supersymmetries (at $x_1$) is greater that the number of global supersymmetries. This is local supersymmetry enhancement.

\vspace{1em}
In a nutshell, the idea of local supersymmetry enhancement is that: Given a set of global supersymmetries, there sometimes exists a whole moduli space of brane/string systems, parametrised by $\{\alpha_i(x) \}$, preserving those same global supersymmetries, but whose number of local supersymmetries is enhanced. The procedure of local supersymmetry enhancement (to an object with 16 local supersymmetries) is a two-step procedure consisting in:
\begin{itemize}
	\item[1.] identify the additional excitations (that we will refer to as ``glues'') to make a bound state;
	\item[2.] determine the charge-to-mass ratios $\{\alpha_i(x) \}$.
\end{itemize}

To our understanding, the presence of a non-vanishing charge-to-mass ratio for a given type of branes/strings indicates the presence of brane/string charges along the bound-state worldvolume, and does not necessarily imply the physical existence the said brane/string wrapping the indicated directions.
When it is the case however, there are requirements in order for the global solution to be consistent. For instance, one has to constrain the $\alpha_i$'s so that that the charge of some brane/strings wrapping some cycles be constant along the cycle. 
Note also that the choice of the glues is not necessarily unique. For example the supersymmetries of NS5$(y1234)$ and F1$(y)$ in Type IIB can be locally enhanced by the pair of glues D5$(y1234)$--D1$(y)$ or KKM$(1234\psi;y)$--P$(\psi)$.

\subsection{Local supersymmetries of the F1-P system}
\label{ssec:F1-P_susies}

Let us go back to our example of F1-P system.
In Section \ref{ssec:LSE_first_ex}, we have identified the dipole charges we need to add in order to locally enhance the supersymmetries of F1$(y)$ and P$(y)$: the ``glues'' we want are of the form F1$(1)$ and P$(1)$, where 1 denotes a direction orthogonal to $y$. 
Let us denote this local supersymmetry enhancement (LSE) by
\be
\begin{bmatrix}
\mathrm{F1}(y) \\
\mathrm{P}(y)
\end{bmatrix} 
\longrightarrow
\( \, \mathrm{F1}(1), \, \mathrm{P}(1) \, \) \,.
\ee
Thus, the candidate projector is of the form:
\be \label{Pi_F1-P}
\hat\Pi_{\mathrm{F1-P}}=\frac12 \left(1 + \alpha_1 P_{\mathrm{F1}(y)}+ \alpha_2 P_{\mathrm{P}(y)}+ \alpha_3 P_{\rm F1(1)}+ \alpha_4 P_{\mathrm{P}(1)}  \right) \,.
\ee

We now wish to determine constraints on the coefficients $\alpha_1, \dots, \alpha_4$. First, we compute $\hat\Pi_{\mathrm{F1-P}}^2$ and impose it to be equal to $\hat\Pi_{\mathrm{F1-P}}$:
\begin{align} \label{Pi_squared_F1-P}
\hat\Pi_{\mathrm{F1-P}}^2=\frac14 \biggl[ 1 + \alpha_1^2 + \alpha_2^2 + \alpha_3^2 + \alpha_4^2 &+ 2\(\alpha_1 P_{\mathrm{F1}(y)}+ \alpha_2 P_{\mathrm{P}(y)} + \alpha_3 P_{\rm F1(1)}+ \alpha_4 P_{\mathrm{P}(1)} \) \nonumber\\
&+ \sum_{i\neq j} \alpha_i\alpha_j P_iP_j  \biggr] \,.
\end{align}
Reading off the coefficients in front of the identity imposes
\begin{equation} \label{cond_sum_coefs}
	\alpha_1^2 + \alpha_2^2 + \alpha_3^2 + \alpha_4^2 ~=~ 1 \,.
\end{equation}
In addition, we also need that the second line of \eqref{Pi_squared_F1-P} vanish. The involutions that anti-commute do not contribute in the sum, while those which commute (here those of F1$(y)$ with P$(y)$, and F1$(1)$ with P$(1)$) lead to the constraint
\begin{equation} \label{cond_projector_coefs}
	\alpha_1 \alpha_2 + \alpha_3 \alpha_4 ~=~ 0 \,.
\end{equation}

Besides, the condition (\ref{eq:global_susy_general}) is written here as:
\begin{align}
	\hat\Pi_{\text{\rm F1-P}} ~=~ 
	 f_1 \Pi_{\mathrm{F1}(y)}+ f_2 \Pi_{\mathrm{P}(y)} \,.
	 \label{eq:F1P_global_form}
\end{align}
One needs to `factorise' the projector in (\ref{Pi_F1-P}) by the involutions of the main excitations, $P_{\mathrm{F1}(y)}$ and $P_{\mathrm{P}(y)}$. We can write for instance
\be \label{choice_factorization1}
P_{\mathrm{F1}(1)} = (-\Gamma^{y1}) \, P_{\mathrm{F1}(y)} \,, \qquad
P_{\mathrm{P}(1)} = (-\Gamma^{y1}) \, P_{\mathrm{P}(y)} \,
\ee
so that $\hat\Pi_{\mathrm{F1-P}}$ in (\ref{Pi_F1-P}) is rewritten as
\begin{align} \label{F1-P_global_intermediate}
\hat\Pi_{\mathrm{F1-P}} ~=~& \(\alpha_1-\alpha_3\Gamma^{y1} \) \Pi_{\mathrm{F1}(y)} + \(\alpha_2-\alpha_4\Gamma^{y1} \) \Pi_{\mathrm{P}(y)} \nonumber\\ 
& \hspace{2em} + \frac{1}{2} - \frac{1}{2}\(\alpha_1-\alpha_3\Gamma^{y1} \) - \frac{1}{2}\(\alpha_2-\alpha_4\Gamma^{y1} \) \,.
\end{align}
In order to satisfy \eqref{eq:F1P_global_form}, one possible solution is
\be
f_1 ~=~ \alpha_1-\alpha_3\Gamma^{y1} \,,\qquad f_2 ~=~ \alpha_2-\alpha_4\Gamma^{y1} \,,
\ee
so that
\be \label{F1-P_constraint_global}
\alpha_1 + \alpha_2 ~=~ 1  \,,\qquad  \alpha_3 + \alpha_4 ~=~ 0 \,.
\ee
The first equation in \eqref{F1-P_constraint_global} is the BPS condition, $Q_1+Q_2=M$, and the second one equates the local charge of the glues.

Note that instead of \eqref{choice_factorization1}, one can also choose to write
\be
P_{\mathrm{P}(1)} = (-\Gamma^{y1}\sigma_3) \, P_{\mathrm{F1}(y)} \,, \qquad
P_{\mathrm{F1}(1)} = (-\Gamma^{y1}\sigma_3) \, P_{\mathrm{P}(y)} \,
\ee
which implies another choice of $f_1$, $f_2$
\be
f_1 ~=~ \alpha_1 - \alpha_4 \Gamma^{y1} \sigma_3 \,,\qquad f_2 ~=~ \alpha_2 - \alpha_3 \Gamma^{y1} \sigma_3 \,,
\ee
but imposes the same constraints \eqref{F1-P_constraint_global}: Although the choice of $f_1$ and $f_2$ is not unique, when one has chosen the consistent glues, the constraints on the charge-to-mass ratios are the same. 

\vspace{1em}
The solution to \eqref{cond_sum_coefs}, \eqref{cond_projector_coefs} and \eqref{F1-P_constraint_global}, assuming that we choose the orientation of space such that the brane charges $Q_1$ and $Q_2$ are positive, can be parameterised by
\begin{align}
\alpha_1= \cos^2\alpha \,, \qquad 
\alpha_2= \sin^2\alpha \,, \qquad 
\alpha_3= \cos\alpha \sin\alpha \,, \qquad 
\alpha_4= -\cos\alpha \sin\alpha \,.
\end{align}
We can thus rewrite \eqref{Pi_F1-P} as:
\begin{align} \label{eq:susy_F1-P}
\hat\Pi_{\mathrm{F1-P}} &= \frac{1}{2} \biggl[ 1+ c P_{\mathrm{F1}(\zeta)} + s P_{\mathrm{P}(\zeta^\perp)} \biggr] \nonumber\\
&= \frac{1}{2} \biggl[ 1+ c \left( c \Gamma^{0y} \sigma_3 + s\Gamma^{01} \sigma_3 \right) + s \left( s \Gamma^{0y} \sigma_3 - c\Gamma^{01} \sigma_3 \right) \biggr] \,,
\end{align}
where $c=\cos \alpha$ and $s=\sin \alpha$.
Equation (\ref{eq:susy_F1-P}) indicates that the centre-of-mass energy of the string is distributed between a piece of string extending along the direction $\zeta$ and its momentum along its orthogonal direction, $\zeta^\perp$.
Therefore, geometrically, the angle $\alpha(y)$ corresponds to the local inclination of the string in the $(y, x_1)$ plane, and depends on the coordinate $y$. See Fig. \ref{fig:F1-P_wiggle}. The projector (\ref{eq:susy_F1-P}) interpolates between that of a pure F1 along $y$ ($\alpha=0$) and that of a pure momentum wave along $y$ ($\alpha=\pm \pi/2$).


\subsection{Microstate geometries of the F1-P black hole}

If one applied the harmonic-function rule to the 1/4-BPS F1-P system in a $\mathbb{R}^{4,1} \times S^1_y \times T^4$ topology, one would find the metric (in string frame) and the dilaton\footnote{For the expression of the Kalb-Ramond field, see e.g. \cite{Mathur:2005zp}.}
\begin{align} \label{metric_BH_F1-P}
ds^2 &= \frac{1}{H_1}  \left[ -dt^2+ dy^2 + \left( H_P -1 \right) (dt+dy)^2 \right] 
+  ds^2_{\mathbb{R}^4} + ds^2_{T^4} \nonumber \\
&= \frac{1}{H_1}  \left[ -du\,dv + \left( H_P -1 \right) dv^2 \right] 
+  ds^2_{\mathbb{R}^4} + ds^2_{T^4}\,, \\
e^{2\phi}&=\frac{1}{H_1} \,,
\end{align}
where we defined the null-like coordinates $u\equiv t+y$, $v\equiv t-y$. The harmonic functions $H_1$ and $H_P$ are of the form in \eqref{eq:sugra_charges_vs_number_branes} (but the supergravity charges, $Q_I$, are expressed differently in terms of the integer charges).
The metric \eqref{metric_BH_F1-P} is that of a black-hole solution in 4+1 dimensions (or black string in 5+1 dimensions), with a horizon at $r=0$.

On the other hand, applying the harmonic-function rule locally to the string carrying transverse oscillations would give a metric and dilaton (see e.g. \cite{Mathur:2005zp} and references therein)
\begin{align} \label{metric_F1-P_bound}
ds^2 &=  \frac{1}{H_1}  \left[ -du\,dv + \left( H_P -1 \right) dv^2 + 2A_i \,dx_i dv \right] 
+  ds^2_{\mathbb{R}^4} + ds^2_{T^4}\,, \\
e^{2\phi}&=\frac{1}{H_1} \,,
\end{align}
which depend on the transverse displacement profile of the string, $\Vec{F}(v)$, in $\mathbb{R}^4$
\begin{align}
H_1 &= 1~+~ \frac{Q_1}{L_T} \int_0^{L_T} \frac{dv}{|\Vec{x}-\Vec{F}(v)|^2} \,, \\
H_P &= 1~+~ \frac{Q_1}{L_T} \int_0^{L_T} \frac{\(\Dot{F}(v)\)^2}{|\Vec{x}-\Vec{F}(v)|^2} dv \,, \\
A_i &= - \frac{Q_1}{L_T} \int_0^{L_T} \frac{\Dot{F}_i(v)}{|\Vec{x}-\Vec{F}(v)|^2} dv \,.
\end{align}
Here the string winds $N_1$ times the circle $S^1_y$, so that its total length is $L_T=2\pi R_y N_1$. The metric is smooth, and instead of a horizon lying at the bottom of an infinite throat for the black hole, such a geometry caps off at the end of a finite throat \cite{Lunin:2001fv,Lunin:2002iz,Mathur:2005zp}. They are denoted as the Lunin-Mathur geometries in the literature.

Of course, such a profile is a classical profile, since the displacement of the string is a classical function, $\Vec{F}(v)$. But where does the `quantumness' come from then? There are two ways to see it. The first way is that the Fourier coefficients of the classical profile $\Vec{F}(v)$ in the non-compact dimensions are actually mapped to the number of \textit{quantum} momentum excitations of the fundamental string (see e.g. \cite{Kanitscheider:2006zf}). 
The second way to see the quantumness is to consider the phase space defined by classical profiles; measuring the volume of this phase space through geometric quantization gives the number of quantum states in the phase space. It turns out this number, once fermionic excitations are taken into account \cite{Taylor:2005db}, matches the entropy of the two-charge black hole \cite{Rychkov:2005ji}:
\be S=2\pi\sqrt{N_1N_P} \,. \ee 

Therefore, although the Lunin-Mathur solutions are characterised by classical profiles which are expected to be produced by a coherent state of string oscillators, the important lesson is that the ``glues'' in the LSE formalism reveal which kind of excitations are relevant at the quantum level in a given background. In the F1-P background, these are the string oscillators, $X^I$; in the D5-D1 frame, the (non-internal) excitations are those of a profile indicating the locus of a Kaluza-Klein monopole in $\mathbb{R}^4$, stabilised by an angular momentum along that profile.

\vspace{1em}
However, the solutions described above are microstate geometries of a two-charge black hole, for which the horizon and the black-hole singularity are located both at $r=0$ (without taking into account higher-curvature corrections) \cite{Dabholkar:2004dq,Sen:2004dp,Hubeny:2004ji}. Therefore, one does not know if the stringy structure resolves the horizon or the singularity at this level.

On the other hand, for three-charge black holes, the horizon and black-hole singularity are separated, even without $\alpha'$ corrections. Understanding whether horizonless geometries can account for the entropy of three-charge black holes will indicate whether the geometries resolve the horizon or the black-hole singularity. Besides, the typical microstates of a two-charge black hole are expected to have a high curvature, and so the supergravity description is expected to fail \cite{Kanitscheider:2007wq}. However, it is possible that this is an artefact of two-charge black holes, due to the proximity of the black-hole singularity with the horizon.

We will thus see how one could apply local supersymmetry enhancement to three-charge brane systems in the following sections.

\subsection{The supertube transition and three-charge systems}

\label{ssec:supertube_transition}

In order to perform the local supersymmetry enhancement (LSE) for three-charge black holes, the microstate geometries has long used the `supertube transition trick' \cite{Bena:2011uw,Mateos:2001qs,Emparan:2001ux}. 
This trick consists in that, for the D1-D5 or D1-D5-P brane system, one can enhance the local supersymmetries of the D1$(y)$ with the D5$(y1234)$ by adding dipole charges of KKM$(1234\psi;y)$\footnote{Here $y$ is the special direction of the Kaluza-Klein monopole.} and P$(\psi)$:
\be
\begin{bmatrix}
\mathrm{D5}(y1234) \\
\mathrm{D1}(y)
\end{bmatrix} 
\longrightarrow
\( \, \mathrm{KKM}(1234\psi;y), \, \mathrm{P}(\psi) \, \) \,,
\ee
where the coordinate $\psi$ parameterises a closed and non self-intersecting curve in the non-compact spatial dimensions, $\mathbb{R}^4$.
The same type of supertube transition can be performed for a NS5-F1 system:
\be
\begin{bmatrix}
\mathrm{NS5}(y1234) \\
\mathrm{F1}(y)
\end{bmatrix} 
\longrightarrow
\( \, \mathrm{KKM}(1234\psi;y), \, \mathrm{P}(\psi) \, \) \,.
\ee

\begin{figure}[h]
\centering
\includegraphics[width=0.5\linewidth]{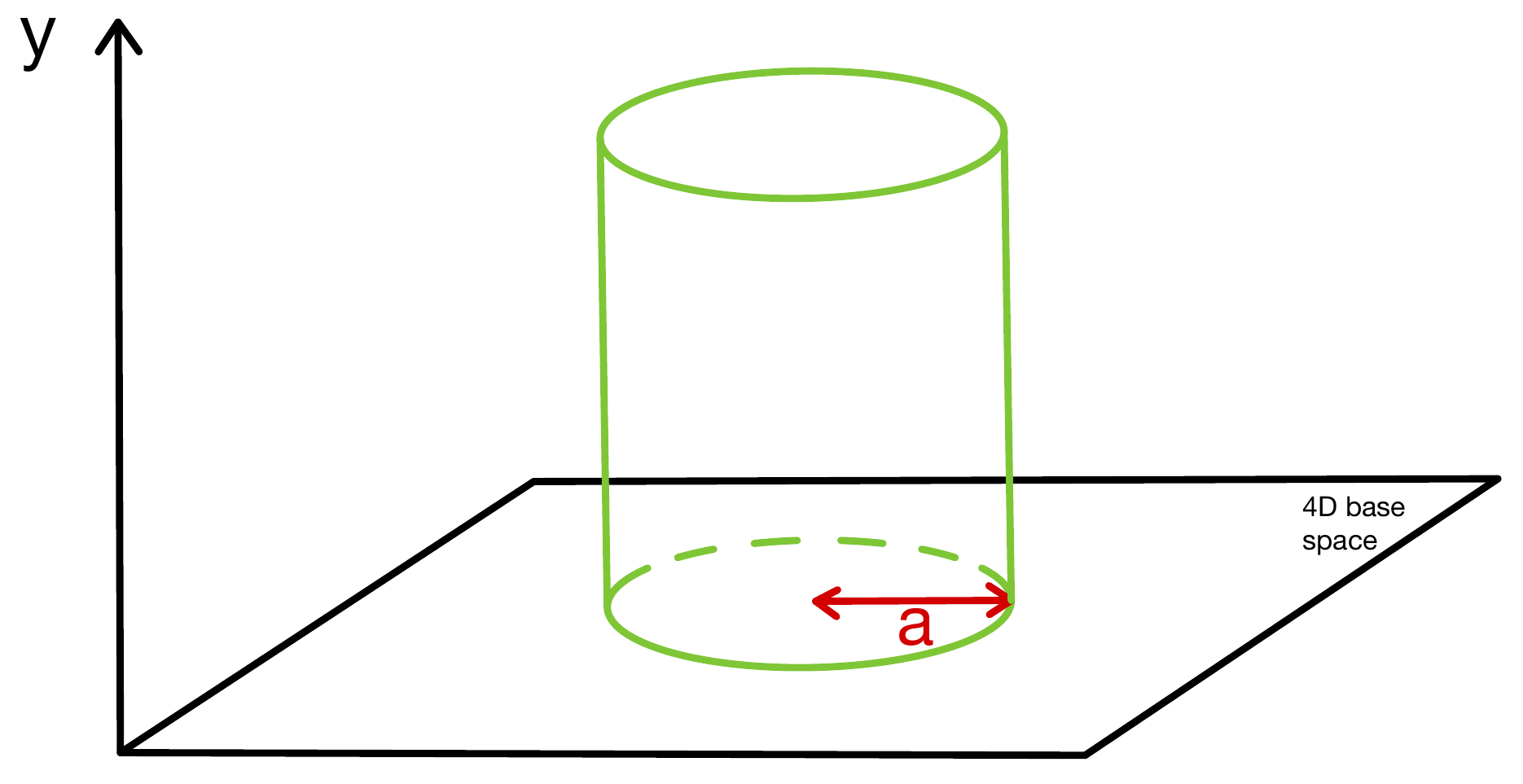}
\caption{Schematic picture of D1-D5 supertube of a circular shape in the base space, here $\mathbb{R}^4$. As one includes the direction of the $y$ circle, the supertube locus corresponds to a cylinder. For the black hole, the brane system lies instead on a straight line along $y$.}
\label{fig:D1-D5_Supertube}
\end{figure}

The key point is that the D1-D5 system (or the F1-NS5 system), through the appearance of the KKM dipole, gains a dimension in the non-compact dimensions; it develops a size, which is stabilised by the presence of the momentum dipole, P$(\psi)$. This trick has the key effect of delocalising the singular brane charges in $\mathbb{R}^4$, from $r=0$ to the supertube locus (the closed curve parametrised by $\psi$). See Fig. \ref{fig:D1-D5_Supertube}. 

This is quite analogous, in electromagnetism, to delocalising a delta-function electric charge in a ring with the same electric charge. In 3+1 dimensions, the electric field sourced by the singular charge is of the form $\frac{Q}{r^2}$; for the ring of radius $R$ with the same electric charge $Q$, the electric field asymptotes to $\sim \frac{Q}{r^2}$ at large $r$, but behaves like $\sim \frac{Q/(2\pi R)}{\Sigma}$ near the ring, where $\Sigma$ is the distance in $\mathbb{R}^3$ to the ring.

Now let us recall that in the computation of the black-hole area at eq. (\ref{area_5_Q1Q5QP}), we used that the harmonic functions behave as $H_i \sim \frac{Q_i}{r^2}$ in the near-horizon region. The presence of the KKM desingularises the brane system which was defined at $r=0$, the metric sourced by the new brane system is less singular; this is a general philosophy of the Fuzzball programme to find horizonless geometries with the same asymptotic mass and charges as the black hole.

Concerning three-charge black holes, one needs to find a way to consistently add the momentum P$(y)$ such that the momentum carriers still have 16 local supersymmetries. Without entering into the details, this has been done in the literature \cite{Bena:2011uw}; and one possibility to have 1/2-BPS momentum carriers is to take the shape modes of the supertube profile along the $S^1_y$ circle. Indeed, locally, the system is a piece of Kaluza-Klein monopole oriented along some direction, boosted by a momentum orthogonal to its surface; it preserves therefore 16 local supersymmetries. However, on the global scale, the 16 local supersymmetries are not the same as one moves along the spatial dimensions of the system, and only 4 supersymmetries are preserved.

Supergravity solutions of a D1-D5-P system blown up by a KKM dipole have been constructed \cite{Bena:2016ypk,Bena:2017xbt}: these are the ``superstrata''. Superstrata approximate the black-hole solution, they have the same charges, asymptotics, and the same near-horizon throat as the black hole. However, instead of the extremal black hole's throat of infinite length, the supertratum's throats caps off smoothly, without the presence of a horizon.

These solutions seem to support the Fuzzball hypothesis \cite{Bena:2022rna}, according to which individual black-hole microstates should differ from the black-hole solution already at the horizon scale (and not only at the scale of the singularity). In particular, there should be a basis of the Hilbert space of black-hole microstates with horizonless microstates.

Nevertheless, it has been shown that the geometric quantization of the superstrata that have been constructed give an entropy that is parametrically smaller than that of the black hole \cite{Shigemori:2019orj,Mayerson:2020acj}:
\be \label{eq:entropy_supertrata}
S_{\textrm{superstrata}} \sim 2\pi \sqrt{N_1N_5}N_P^{1/4} \,,
\ee
which is parametrically smaller than the black-hole entropy, $S\sim 2\pi \sqrt{N_1N_5N_P}$.%
\footnote{Note, however, that the above formula (\ref{eq:entropy_supertrata}) counts the number of superstrata that have been \textit{constructed} in supergravity with \cite{Bena:2016ypk,Bena:2017xbt}, and not the shape-mode superstrata predicted by \cite{Bena:2011uw}, which could be in principle more numerous \cite{Bena:2014qxa}. See also \cite{Li:2022apf}.}

Besides, the superstrata have a non-vanishing angular momentum in $\mathbb{R}^4$ through the appearance of the KKM and angular momentum dipoles, whereas the black-hole solution (the pure D1-D5-P system) has no such angular momentum. If the black-hole horizon's spherical symmetry is broken by a KKM whose supertube locus is at a macroscopic scale, this could be a sign that the superstrata are atypical microstates (in the grand-canonical ensemble) of the D1-D5-P black hole (or, they could be microstates of another black hole, with non-vanishing angular momentum in $\mathbb{R}^4$). (See also \cite{Lin:2022rzw} for discussion about exact spherical symmetry.)

One may argue that, similar to what happens in \cite{Bena:2006kb,Li:2021gbg,Li:2021utg}, once one takes into account the gravitational backreaction, making the supertube profile smaller and smaller in \textit{coordinate space} in $\mathbb{R}^4$ corresponds to making the superstratum's throat deeper and deeper, while the geometry of the cap remains fixed, and does not create singularity at the supergravity level. And thus the KKM and corresponding angular momentum can be arbitrarily small, so that the superstrata are more and more typical microstates. One could be worried, however, that the microstate then degenerates into a black hole; but the quantization of the angular momentum prevents the throat to become infinitely deep and the superstratum to degenerate into the black-hole solution. Nevertheless, one could expect in this case that the quantum corrections over the supergravity solutions to be too large, in the same spirit of \cite{deBoer:2008zn,Li:2021gbg}.

A possible way out of this problem has been studied in \cite{Bena:2022sge}. We studied the possibility that the deep-throat limit of superstrata reaches actually solutions with internal momentum carriers, in the NS5-F1-P frame in Type IIA. Nevertheless, the momentum carriers, consisting of D0-D4 density modes inside the NS5 and moving along the $y$ direction, preserve only 8 supersymmetries, instead of 16. This particular feature implied that their supergravity description exhibits a horizon of zero area (before higher-order corrections). Therefore, they cannot be pure black-hole microstates. In the next section, we will construct momentum carriers with 16 local supersymmetries, as pure microstates of the NS5-F1-P (IIA) or M5-M2-P black hole.

\section{Local supersymmetries with internal excitations}
\label{sec:internal_excitations}

In this Section, we take a different path from the historical approach that lead to the discovery of most fuzzball solutions, like the superstrata. Namely, our aim will be to find microstates with 16 local supersymmetries with pure internal excitations. 
We focus on the NS5-F1-P system. In the next subsection, we first identify the local supersymmetry enhancement (LSE) between the 1/4-BPS F1-P, NS5-P and F1-NS5 systems.

\subsection{Two-charge systems}

The local supersymmetric enhancement of the F1-P system has already been done in Sec. \ref{ssec:F1-P_susies}:
\be
\begin{bmatrix}
\mathrm{F1}(y) \\
\mathrm{P}(y)
\end{bmatrix} 
\longrightarrow
\( \, \mathrm{F1}(1), \, \mathrm{P}(1) \, \) \,.
\ee
Here since we want the excitation to be purely internal, we choose the coordinate $x_1$ to be a coordinate of $T^4$.

\vspace{1em}
The same exercise can be done for the NS5-P system in type IIA. We start with NS5 branes extending along the directions $y, x_1, \dots,x_4$, and momentum along $y$. The involutions associated to them are:
\begin{equation}
	P_{{\rm NS5}(y1234)} ~=~ \Gamma^{0y1234} \,,\qquad P_{\mathrm{P}(y)} ~=~ \Gamma^{0y}\,.
\end{equation}
These involutions commute, and the configuration preserves 8 supersymmetries. What are the glues to make a configuration with 16 local supersymmetries? Like the F1-P system, one could consider NS5$(\psi)$ and P$(\psi)$ as glues, where $\psi$ is a coordinate in $\mathbb{R}^4$. This corresponds to bending the NS5 in the transverse directions and let it wiggle. Contrary to the fundamental string, the NS5-brane does not need to bend in the transverse directions to carry momentum. To make a bound state, one other possibility is to use internal dipolar D4 branes (extending along the directions $x_1,\dots,x_4$) and D0 branes  \cite{Bena:2022sge}:
\be
\begin{bmatrix}
\mathrm{NS5}(y1234) \\
\mathrm{P}(y)
\end{bmatrix} 
\longrightarrow
\( \, \mathrm{D4}(1234), \, \mathrm{D0} \, \) \,.
\ee
The advantage of using the D4-D0 glue is that the momentum carriers are purely internal excitations, i.e. point-like in the non-compact spatial dimensions.
The candidate projector is of the form
\begin{align}
	\hat \Pi_{\text{NS5-P}} ~&=~  \frac12 \(1 + \alpha_1 P_{{\rm NS5}(y1234)} + \alpha_2 P_{\mathrm{P}(y)}+ \alpha_3 P_{\rm D4(1234)}+ \alpha_4 P_{\rm D0} \) \label{eq:NS5P_local_form}
	\\
	&=~ f_1 \Pi_{{\rm NS5}(y1234)} + f_2 \Pi_{\mathrm{P}(y)} \,, \label{eq:NS5P_global_form}
\end{align}
and the solution is found to be of the form
\begin{align}
	\hat \Pi_{\text{NS5-P}} ~=~ \frac12 \[1 + c^2 P_{{\rm NS5}(y1234)} + s^2 P_{\mathrm{P}(y)} - cs P_{\rm D4(1234)} + cs P_{\rm D0} \] \,.
\end{align}
Here $c=\cos \alpha$ and $s=\sin \alpha$. The bound state can be understood geometrically from the M-theory perspective, as a M5 brane along $(y1234)$ with transverse oscillations along the M-theory direction, $x^{11} \equiv z$. The angle $\alpha$ is then the angle that the M5 brane makes locally with the $y$ direction in the $(y,z)$ plane.

\begin{figure}[h]
\begin{center}
\includegraphics[scale=1]{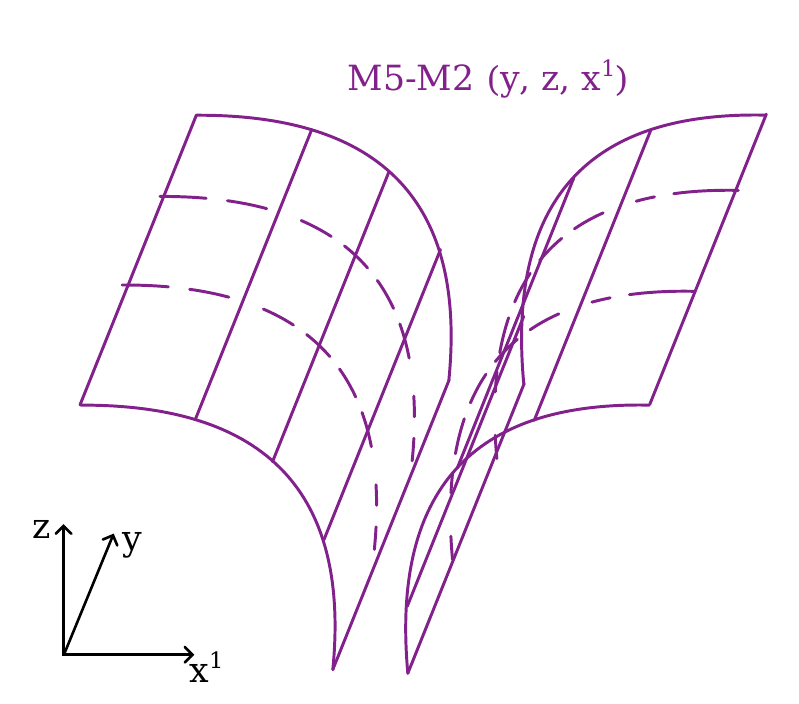}
\caption{  The backreaction of the M5-M2 bound state, projected onto the space $(y,x_1, z)$. The M2-branes pull the M5-branes, forming a furrow. The mechanism is similar to the formation of a Callan-Madacena spike in the D3-F1 brane system. }
\label{fig:M5-M2_furrow}
\end{center}
\end{figure}

\vspace{1em}
Concerning the combination between F1 and NS5, a choice of glues that are purely internal excitations is:
\be
\begin{bmatrix}
\mathrm{NS5}(y1234) \\
\mathrm{F1}(y)
\end{bmatrix} 
\longrightarrow
\( \, \mathrm{D4}(y234), \, \mathrm{D2}(y1) \, \) \,.
\ee
The projector is found to be:
\be \label{projector_NS5-F1}
\hat \Pi_{\text{NS5-F1}} ~=~ \frac12 \[1 + c^2 P_{{\rm NS5}(y1234)} + s^2 P_{\mathrm{F1}(y)}+ cs P_{\mathrm{D4}(y234)} + cs P_{\mathrm{D2}(y1)} \] \,,
\ee
with $c=\cos \beta$ and $s=\sin \beta$, and $\beta$ depends in principle on the coordinates $(y,1,2,3,4)$. The uplift in M theory of the projector \eqref{projector_NS5-F1} is
\begin{align} \label{eq:projector_M5-M2}
\Pi_{\mathrm{M5-M2}} &= \frac{1}{2} \[ 1 
+  c^2 P_{\mathrm{M5}(y1234)} + s^2 P_{\mathrm{M2}(yz)} - c s P_{\mathrm{M5}(y234z)} + cs P_{\mathrm{M2}(y1)} \] \nonumber \\
&= \frac{1}{2} \[ 1 
+  c \( c P_{\mathrm{M5}(y1234)} + s P_{\mathrm{M5}(yz234)} \) + s \( c P_{\mathrm{M2}(y1)} + s P_{\mathrm{M2}(yz)} \) \] \,.
\end{align}
The bound state can be understood geometrically from the M-theory perspective. The original brane system consists of a M2 brane extended along $(y,z)$ ending on a M5 brane extended along $(y,1,2,3,4)$. The M2 brane pulls the M5 brane towards the direction it extends, $z$; this is similar to the Callan-Maldacena effect \cite{Callan:1997kz}, where a F1 ending orthogonally on a D3 brane pulls the pulls the D3 brane towards itself. This pulling effect happens along all the M5-M2 common direction, $y$, and so the bound state looks like a furrow along $y$. See Fig. \ref{fig:M5-M2_furrow}.
The angle $\beta$ in \eqref{eq:projector_M5-M2} interpolates between the projector of a pure M5 brane along $(y1234)$ and that of a M2 brane along $(yz)$. Geometrically, $\beta$ is the angle the M2-M5 bound state makes with the $x^1$ direction in the $(x^1,z)$ plane. As the angle of the bending, $\beta$, increases, the charge of the M2 increases as well.
The interpolation between the M5-brane charge and the M2-brane charge is reminiscent of dyonic M2/M5 brane \cite{Izquierdo:1995ms}. But here, the proportion between the M2 and M5 charges depends also on the orientation, $\beta$, of the piece of the furrow. The best-fitted supergravity solution to this system is perhaps given in \cite{Lunin:2007mj}.

\subsection{The projector for binding NS5, F1 and P all together}

In the previous subsection, we have worked out the local supersymmetry enhancement (LSE) in order to construct bound states of the F1-P, NS5-P and NS5-F1 systems: In particular, we have determined the pairs of glues (or dipoles) one needs to add to the system to complete the LSE. Such bound states involve only dipoles that exhibit a spherical symmetry in the non-compact spatial dimension. In order to make a bound state of the F1-NS5-P system, an intuitive guess is to combine all the two-by-two glues together, and check if the projector conditions are satisfied. The list of dipoles is summarised in Table \ref{tab:NS5-F1-P_and_glue}.
\begin{table}[h]
    \centering
\begin{tabular}{|c|c|c||c|c|c|c|c|c|}
\hline
NS5$(y1234)$ & F1$(y)$      & P$(y)$       & D4$(y234)$ & D2$(y1)$ & D4$(1234)$ & D0       & F$(1)$   & P$(1)$   \\ \hline
$\bigotimes$ & $\bigotimes$ &              & $\times$   & $\times$ &            &          &          &          \\ \hline
$\bigotimes$ &              & $\bigotimes$ &            &          & $\times$   & $\times$ &          &          \\ \hline
             & $\bigotimes$ & $\bigotimes$ &            &          &            &          & $\times$ & $\times$ \\ \hline
\end{tabular}
    \caption{Each line describes a two-charge bound state whose charges are two of the three charges of the NS5-F1-P brane systems (denoted by $\bigotimes$). Each bound state contains two more dipole charges, denoted by $\times$. We attempt to construct a three-charge bound state with NS5-F1-P and all six dipole charges.}
    \label{tab:NS5-F1-P_and_glue}
\end{table}
The projector for such bound state is then written in the form
\begin{align}
 \hat \Pi_{\text{NS5-F1-P}} ~=~ 
 	\begin{aligned}[t] 
	\frac{1}{2} \biggl[ 1 &+ \alpha_1 P_{\mathrm{NS5}(y1234)} + \alpha_2 P_{\mathrm{F1}(y)} + \alpha_3 P_{\mathrm{P}(y)}\\
        &+ \alpha_4 P_{\mathrm{D4}(y234)}
	 +\alpha_5 P_{\mathrm{D2}(y1)} +\alpha_6 P_{\mathrm{P}(1)} +\alpha_7 P_{\mathrm{F1}(1)} +\alpha_8 P_{\mathrm{D4}(1234)} +\alpha_9 P_{\mathrm{D0}} \biggr] \,.
	\end{aligned} \label{eq:NS5F1P_local_projector}
\end{align}
In addition to the condition that $\hat \Pi_{\text{NS5-F1-P}}$ should be a projector, we further impose $\hat \Pi_{\text{NS5-F1-P}}$ to be written of the form \eqref{eq:global_susy_general}:
\begin{align}
\hat \Pi_{\text{NS5-F1-P}} =~ f_1 \Pi_{\mathrm{NS5}(y1234)}+ f_2 \Pi_{\mathrm{F1}(y)}+ f_3 \Pi_{\mathrm{P}(y)}\,. \label{eq:NS5F1P_global_projector}
\end{align}
The solution to these equations is shown to be parameterised by three real numbers $(a,b,c)$ satisfying the constraint $a^2 + b^2 + c^2 = 1$:
\begin{equation}
\hat \Pi_{\text{NS5-F1-P}} ~= 
	\begin{aligned}[t]
		\frac{1}{2} \biggl[1 &+ a^2 P_{\mathrm{NS5}(y1234)} + b^2 P_{\mathrm{F1}(y)} + c^2 P_{\mathrm{P}(y)}\\
        &+ ab \left( P_{\mathrm{D4}(y234)} + P_{\mathrm{D2}(y1)} \right)
		 + bc \left( P_{\mathrm{P}(1)} - P_{\mathrm{F1}(1)} \right)
		- ac \left( P_{\mathrm{D4}(1234)} - P_{\mathrm{D0}} \right) \biggr] \,.
	\end{aligned}
	\label{eq:projector_NS5-F1-P_bound_final}
\end{equation}

In terms of M-theory ingredients, the projector is written as:
\begin{align} \label{eq:projector_M5-M2-P}
\hat\Pi_{\mathrm{NS5-F1-P}}= \frac{1}{2} \biggl[ 1 
+  a \hat P_{\mathrm{M5}}
+  b \hat P_{\mathrm{M2}}
+  c \hat P_{\mathrm{P}}  \biggr] \,,
\end{align}
where
\begin{align} 
\hat P_{\mathrm{M5}} &\equiv a P_{\mathrm{M5}(y1 \,234)} + b P_{\mathrm{M5}(yz\,234)} + c P_{\mathrm{M5}(z1\, 234)} \label{eq:projector_for_M5} \,,\\
\hat P_{\mathrm{M2}} &\equiv  a P_{\mathrm{M2}(y1)} + b P_{\mathrm{M2}(y\,z)} + c P_{\mathrm{M2}(z1)} \label{eq:projector_for_M2} \,,\\
\hat P_{\mathrm{P}} &\equiv  a P_{\mathrm{P}(z)} + b P_{\mathrm{P}(1)} + c P_{\mathrm{P}(y)} \,, \label{eq:projector_for_P}
\end{align}
and the brane involutions are given in Appendix \ref{sec:projectors_and_involutions_for_branes}.

The (local) geometric interpretation of the bound state is as follows. We can span the $(y, z, 1)$ space using orthonormal vectors $(u_{y},u_z,u_1)$.
 Let $u^\perp_{\mathrm{M5}}$ be the unit vector orthogonal to the two-dimensional M5-brane surface in the $(y, z, 1)$ space. Let $u^\perp_{\mathrm{M2}}$ be its equivalent for the M2-brane, and $u_{\mathrm{P}}$ the unit vector along the direction of the momentum P.
Then, by choosing the orientation signs appropriately, one can show  that the equations (\ref{eq:projector_for_M5}), (\ref{eq:projector_for_M2}) and (\ref{eq:projector_for_P}) imply successively
\begin{align}
a &= u^\perp_{\mathrm{M5}} \cdot u_{z} \,, \qquad b = u^\perp_{\mathrm{M5}} \cdot u_{1} \,, \qquad c = u^\perp_{\mathrm{M5}} \cdot u_{y} \,,\\
a &= u^\perp_{\mathrm{M2}} \cdot u_{z} \,, \qquad b = u^\perp_{\mathrm{M2}} \cdot u_{1} \,, \qquad c = u^\perp_{\mathrm{M2}} \cdot u_{y} \,,\\
a &= u_{\mathrm{P}} \cdot u_{z} \,,\, \, \qquad \, b = u_{\mathrm{P}} \cdot u_{1} \,,\,\, \qquad \, c = u_{\mathrm{P}} \cdot u_{y} \,.
\end{align}
Hence, these equations simply imply that:
\be \label{unit_vectors_M5M2P}
u^\perp_{\mathrm{M5}} = u^\perp_{\mathrm{M2}} =  u_{\mathrm{P}} \; = a \, u_{z} + b \, u_1 + c \, u_y \,.
\ee
Therefore, the M5 brane is parallel to the M2 brane, and the momentum P is orthogonal to both of them. This is why there are locally 16 supersymmetries.

Thus, the coefficients $(a,b,c)$ have two physical roles. On the one hand, equation \eqref{eq:projector_M5-M2-P} tell that the bound state is a mix of M5, M2 and P along some directions, and $(a,b,c)$ specify the relative charge between each ingredient (M5, M2 and P) in the mix. On the other hand, in \eqref{unit_vectors_M5M2P}, $(a,b,c)$ specify the orientation of the local piece of the M5-M2-P bound state, made of a parallel M5-M2 that is boosted orthogonally along its surface. In other words, the local orientation of the bound state determines the relative charges and vice-versa.

\begin{figure}[h]
\begin{center}
\includegraphics[scale=1]{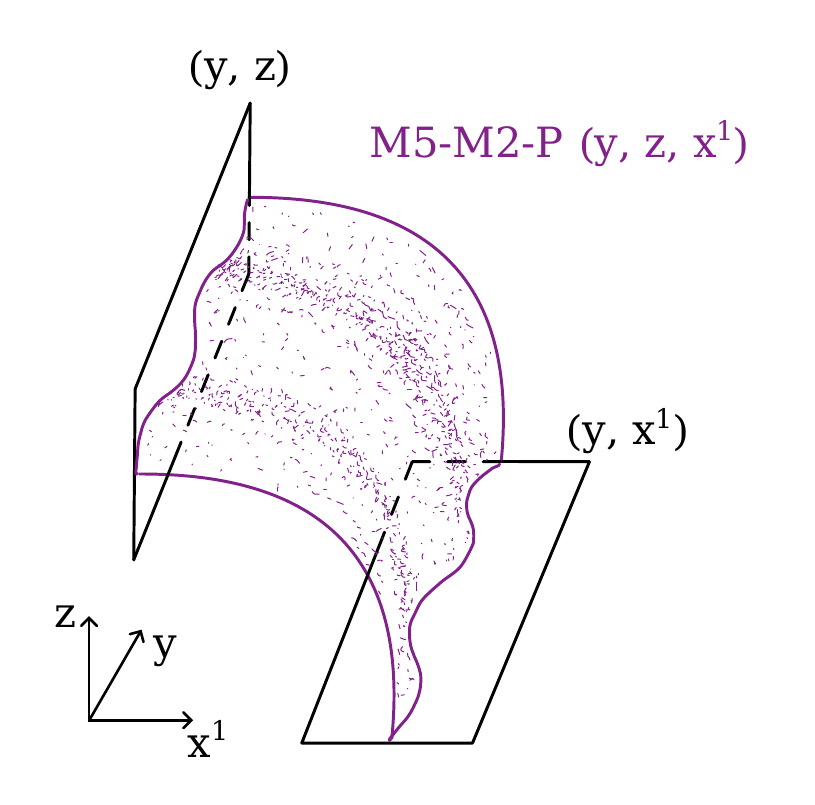}
\caption{Schematic picture of the M5-M2-P bound state, at the scale of the M5-M2 furrow.}
\label{fig:M5-M2-P_wiggle}
\end{center}
\end{figure}

At a larger scale, the M5-M2-P bound state consists of a piece of M5-M2 furrow (of the previous subsection) carrying momentum along $y$ by having oscillations transverse to its surface. See Fig. \ref{fig:M5-M2-P_wiggle}.

When $a=0$, the M5-M2-P system becomes an M2-P system. The M2 wraps the direction $z$ and carries a momentum wave  in the $y$ direction, which causes it to wiggles in the $(y,1)$ plane. Locally, the slope of the wiggling M2 brane is parameterized by an angle, $\alpha$, that the M2 brane in  $(y,1)$ plane the makes with the $y$ direction.

When $b=0$, the M5-M2-P system becomes an M5-P system. The M5 wraps the directions $(1,2,3,4)$ and a one-dimensional line profile in the $(y,z)$ plane. The M5 carries a momentum wave  in the $y$ direction, which causes the line profile to wiggles along a linear profile in the $(y,z)$ plane. Locally, the slope of the wiggling M5 brane is parameterized by an angle, $\alpha$, that the M5 brane line in $(y,z)$ plane the makes with the $y$ direction.

Finally, when $c=0$, M5-M2-P system becomes an M5-M2 system. The brane configuration describes a piece of an M2-M5 furrow that is independent on $y$ and has no momentum in the $y$ direction. As we explained in the previous subsection, this furrow comes from the Callan-Maldacena-like interaction \cite{Callan:1997kz} between a `horizontal' M5-brane wrapping $(y,1)$ and a `vertical' M2-brane wrapping $(y,z)$. Locally, a piece of this furrow is characterized by an angle $\beta$ with respect to the direction 1 in the $(1,z)$ plane.

Generically, when neither of $a$, $b$, or $c$ vanish, the M5-M2-P system describes a piece of the two-dimensional M5-M2 furrow surface, making an angle $\beta$ with the direction 1 in the $(1,z)$ plane, wiggling along its orthogonal direction in the $(1,z)$ plane, and thus carrying momentum along the $y$ direction.

\subsection{The supermaze}

In the previous subsection, we described how the geometry of the M5-M2-P bound state at the most local scale (eq. \eqref{unit_vectors_M5M2P}) and at the scale of the M5-M2 furrow (Fig. \ref{fig:M5-M2_furrow}). What happens at the scale of the entire microstate, which extends on the whole $T^4$ and M-theory circle, $S_z^1$?

\begin{figure}[h]
\begin{center}
 \includegraphics[width=.7\linewidth]{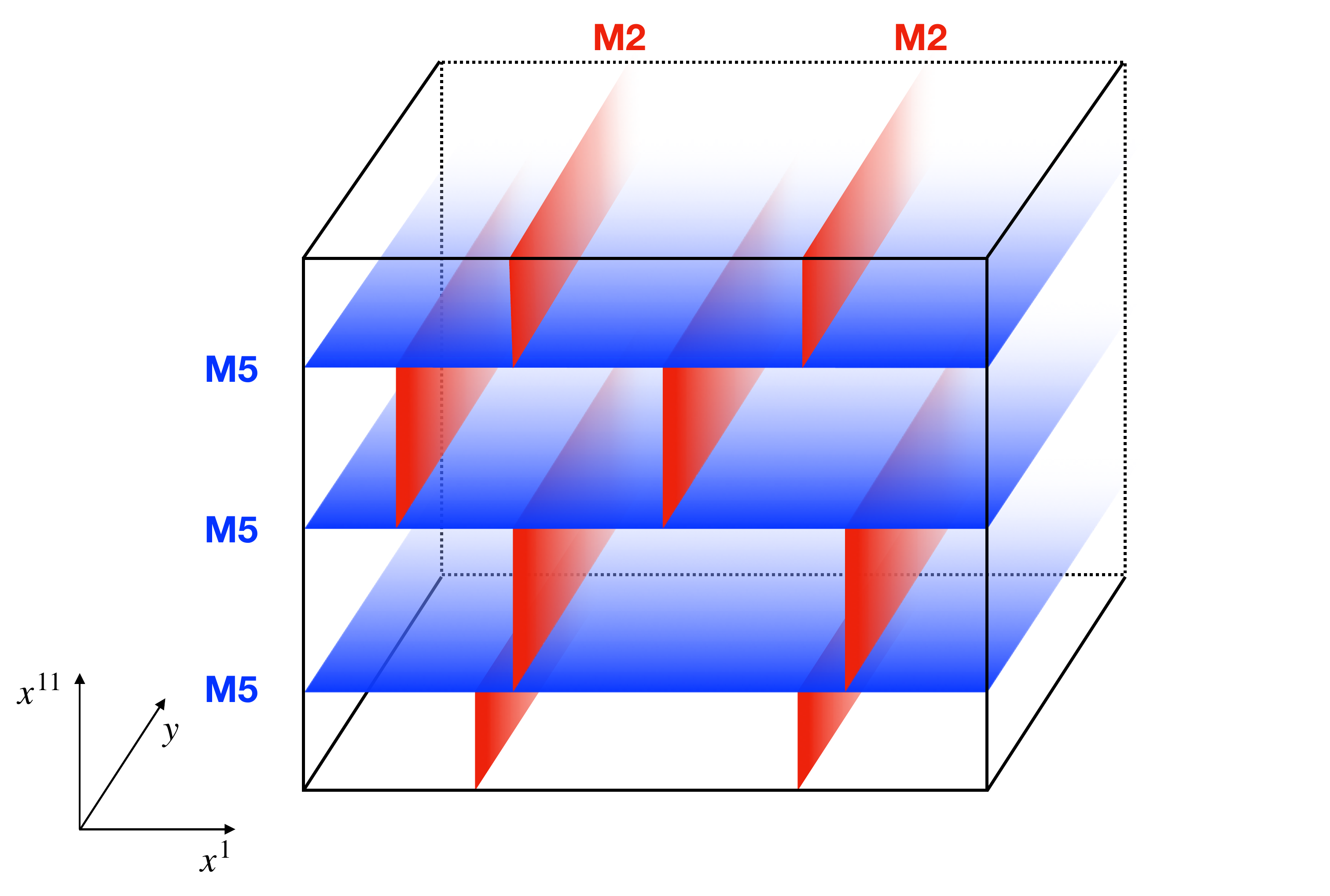}
\caption{Cross-section of $N_1=2$ M2 branes splitting into strips between $N_5=3$ M5 branes. The vertical axis is the M-theory direction, and the horizontal axis represents one of the internal directions of the M5 branes, $x^1$. The strips can carry momentum along the $y$-circle, which is common to the M2 and M5 branes.}
\label{fig:M2_fractionated_M5}
\end{center}
\end{figure}

For a M5-M2 system that is made of $N_5$ M5 branes wrapping $(y1234)$ and $N_1$ M2 branes wrapping $(yz)$, there is no force separating the M2 branes into $N_1N_5$ strips of M2 branes ending on the M5 branes, see Fig. \ref{fig:M2_fractionated_M5}.
These $N_1N_5$ strips of M2 branes are numerous, so the dimension of the moduli space associated to their degrees of freedom is large. For the M5-M2-P black hole, these strips can carry momentum by their transverse oscillations in the $T^4$. Each strip carries 4 bosonic degrees of freedom (for the 4 spatial dimensions in $T^4$), so overall the system has $4 N_1 N_5$ independent bosonic degrees of freedom. 

Reducing to Type IIA, the M2-M5-P black hole becomes a F1-NS5-P black hole. And the fractionated M2 strips become little strings, fractionated by the presence of the NS5 branes and living in their worldvolume \cite{Seiberg:1997zk,Kutasov:2001uf}. Again, they are numerous and act as $N_1 \times N_5$ independent momentum carriers. In the Cardy limit, the entropy of these oscillations and of their fermionic superpartners is $S_{\rm little~strings} = 2 \pi \sqrt{(4+2) {N_1 N_5 N_p \over 6}}$, reproducing precisely the entropy of the F1-NS5-P black hole. We will refer to these microstates as the ``Dijkgraaf-Verlinde-Verlinde-Maldacena (DVVM) microstates'' \cite{Dijkgraaf:1996cv,Maldacena:1996ya} or ``little-string microstates."

\begin{figure}[h]
	\centering
	\includegraphics[width=.31 \textwidth ]{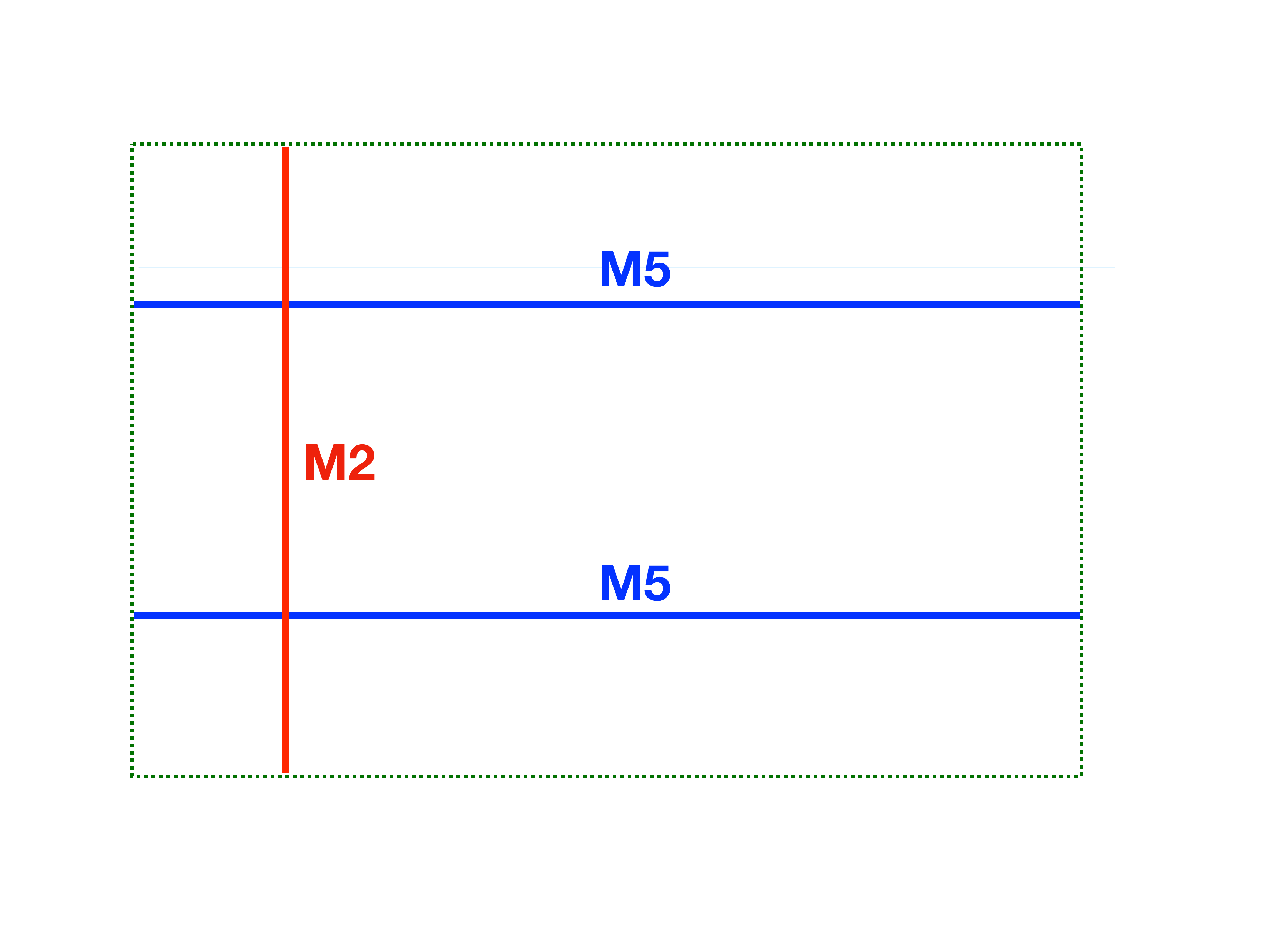}
	\includegraphics[width=.31 \textwidth]{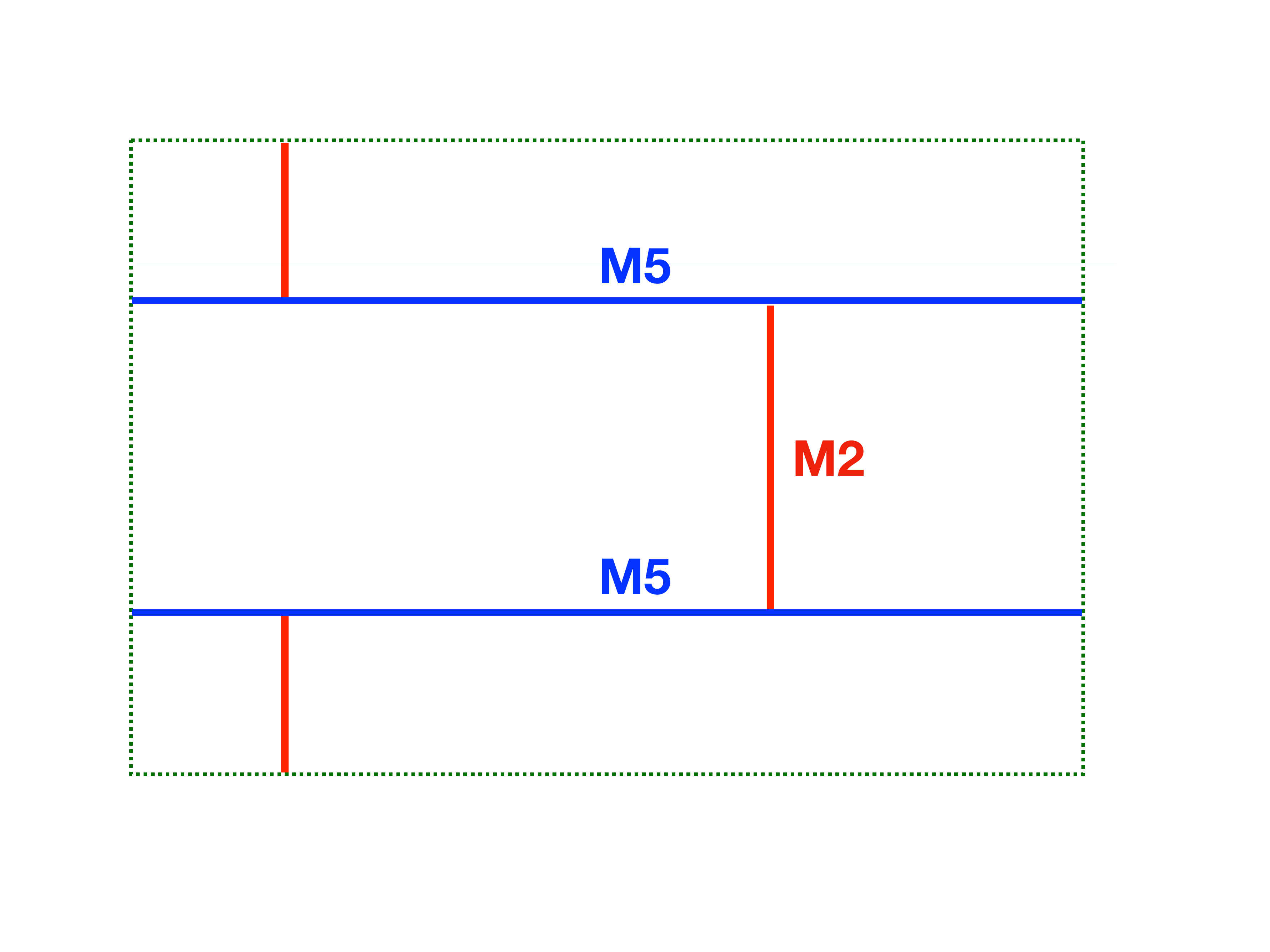}
	\includegraphics[width=.31 \textwidth]{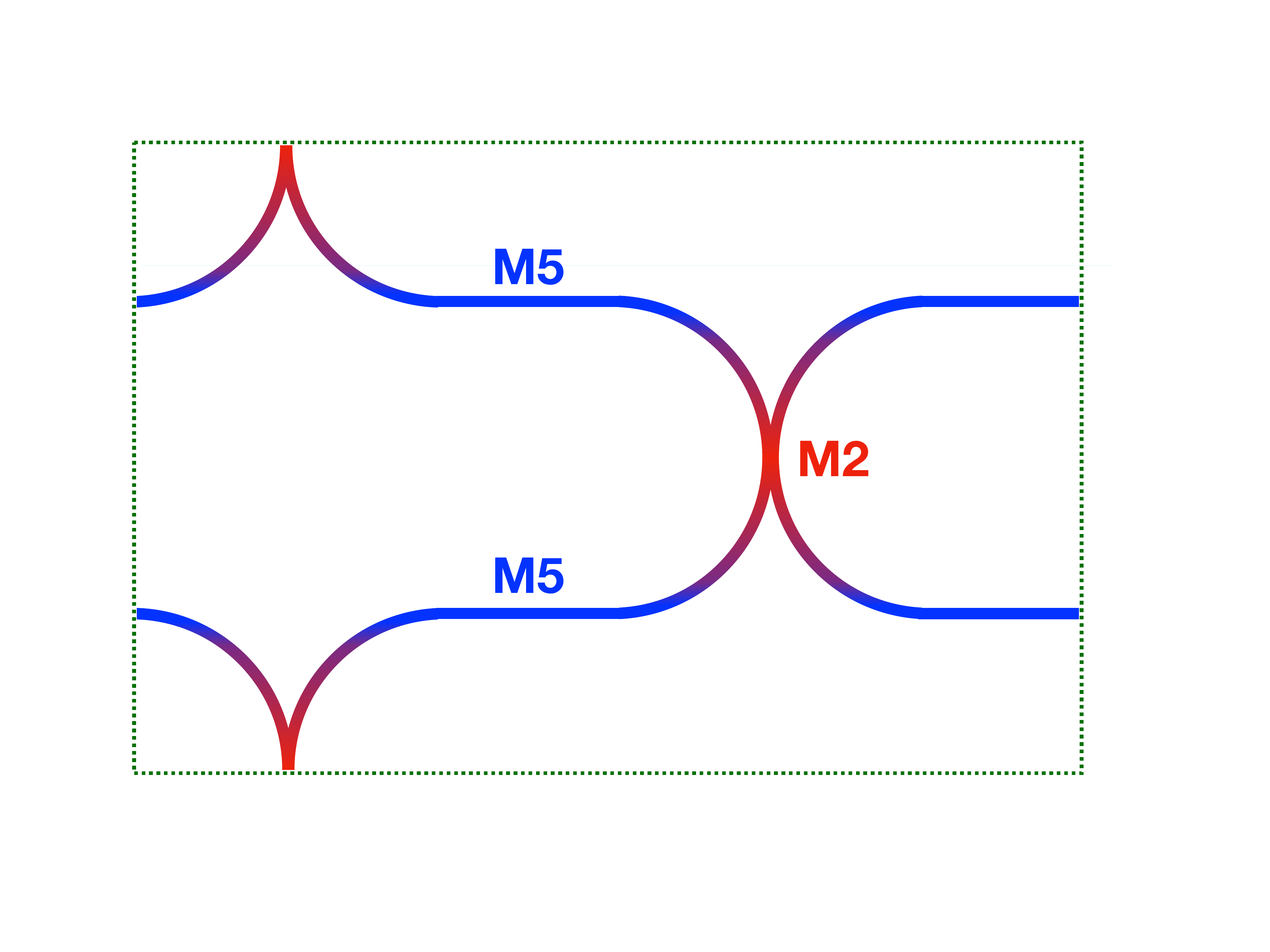}
	\caption{The fractionation of M2 branes into strips and the super-maze: Before the fractionation  (left panel) the M2 brane does not interact with the M5 branes, and can be freely taken away. After the fractionation, each strip of the M2 branes can move independently, giving the na\"ive configuration in the middle panel. However, the M2 strips pull on the M5 brane, creating the supermaze depicted in the right panel.	}
	\label{fig:flux_before_after1}
\end{figure}

The results in the previous subsection about local supersymmetry enhancement of the M2-M5-P bound state tell about another perspective on the DVVM microstates and how they carry momentum. The local transition from a pair of orthogonal M2-M5 branes into a M2-M5 furrow (Fig. \ref{fig:M5-M2_furrow} implies that at a global scale, a DVVM microstate transitions into a kind of labyrinth, or maze, that we call `supermaze'. See Fig. \ref{fig:flux_before_after1}.

Now, let us look back to the projector \eqref{eq:projector_NS5-F1-P_bound_final} and \eqref{eq:projector_M5-M2-P}.

Because of the constraint $a^2+b^2+c^2=1$, there are three obvious ways to parameterise the coefficients $(a,b,c)$ in terms of the M5-M2 bending angle, $\beta$, and of the wiggle angle, $\alpha$. 

A first solution is 
\be \label{M2_mode}
\begin{pmatrix} a \\ b \\ c \end{pmatrix}
=\begin{pmatrix} \cos \beta \\ \cos \alpha \sin \beta \\ \sin \alpha \sin \beta \end{pmatrix}\,.
\ee
This solution corresponds to the furrow generalisation of a M2 strip carrying momentum through its transverse motion, in the four-torus, represented here by the direction of $x^1$. The angle $\beta$ interpolates between an immobile M5-brane ($\beta=0$) and a wiggling M2-strip ($\beta=\pi/2$).

A second solution is 
\be 
\begin{pmatrix} a \\ b \\ c \end{pmatrix}
=\begin{pmatrix} \cos \alpha \cos \beta \\ \sin \beta \\ \sin \alpha \cos \beta \end{pmatrix}\,.
\ee
This solution corresponds to the furrow generalisation of a M5-brane carrying momentum through its transverse motion, in the eleventh dimension, $z$. The angle $\beta$ interpolates between a wiggling M5-brane ($\beta=0$) and an immobile M2-strip ($\beta=\pi/2$).

A third solution is
\be \label{coupled_M2M5_mode}
\begin{pmatrix} a \\ b \\ c \end{pmatrix}
=\begin{pmatrix} \cos \alpha \cos \beta \\ \cos \alpha \sin \beta \\ \sin \alpha \end{pmatrix}\,.
\ee
This solution corresponds to coupling of the M5-motion mode and the M2-motion mode. The angle $\beta$ interpolates between a wiggling M5-brane ($\beta=0$) and a wiggling M2-strip ($\beta=\pi/2$). This is momentum mode that is represented in Figure \ref{fig:M5-M2-P_wiggle}.

As the entropy of the NS5-F1-P black hole to come from the fractionated M2-strips, one expects that the mode (\ref{M2_mode}) is the most entropic mode.

\section{Conclusion}
\label{sec:conclusion}


In string theory, the historical answer to the question ``What are the black-hole microstates?'' has been, in the D1-D5-P black hole context, to specify, in the open-string regime ($g_sN\ll 1$), what the momentum carriers are and how the momentum is distributed. For the D1-D5-P black hole, the fact that the momentum is mostly carried by the 1-5 open strings (in the regime $g_sN\ll 1$) make it difficult to track these degrees of freedom in a regime where $g_s$ becomes bigger, all the way to the supergravity scale ($g_sN\gg 1$, $g_s\ll 1$). 

But in the F1-NS5-P frame, the momentum is known to be carried by the transverse motion (in $T^4$) of the fractionated/little strings; equivalently, for the M2-M5-P black hole, the bosonic part of the entropy (which contributes to $\sqrt{\frac{4}{6}}$ of the total entropy of the black hole) comes from the transverse motion in $T^4$ of the $N_2N_5$ fractionated M2 strips between the M5 branes -- in this proceeding we called them the Dijkgraaf-Verlinde-Verlinde-Maldacena (DVVM) microstates \cite{Dijkgraaf:1996cv,Maldacena:1996ya}. 
We argue that these momentum carriers admit a natural local supersymmetry enhancement (LSE), and that at the scale of the $T^4$, the DVVM microstates backreact into a maze-like brane system, dubbed the \textit{supermaze}. The LSE of M2 and M5 corresponds to a Callan-Maldacena-spike (or BIon) effect \cite{Callan:1997kz} to the M2 strips ending on M5 branes, forming a ``furrow'' along the common M2-M5 direction. Then, the simultaneous LSE of M2, M5 and P corresponds, for the M2-M5 furrow, to carry momentum by having ripples that are orthogonal to its surface. The amplitudes of the ripples can be turned off at the loci where the M5 charge is maximal (and the M2 charge is vanishing), and turned on when the M2 charge non-vanishing. Such a form of the ripples corresponds to the motion of the M2 strips alone (in the regime where the branes were not interacting): Therefore, the supermazes account for $\sqrt{\frac{4}{6}}$ of the total entropy of the black hole as well.

Historically, the Microstate Geometries programme has endeavoured to find supergravity solutions whose corresponding brane description is based on the desingularization of the D1-D5-P brane system into a KKM-P dipole, and where the momentum is carried by the shape modes of this KKM-P dipole (see Section \ref{ssec:supertube_transition}). These momentum carriers may lead to atypical microstates of the D1-D5-P or F1-NS5-P black hole. And instead, we have found for the first time microstates with 16 local supersymmetries which carry the momentum through their motion in the internal/compact dimension \cite{Bena:2022wpl}.

The LSE of the DVVM microstates of the F1-NS5-P black hole reveal how the microstates behave as one tunes the string coupling constant, $g_s$, to greater values and the branes start interacting.

\begin{figure}[h]
	\centering
	\includegraphics[height=100px]{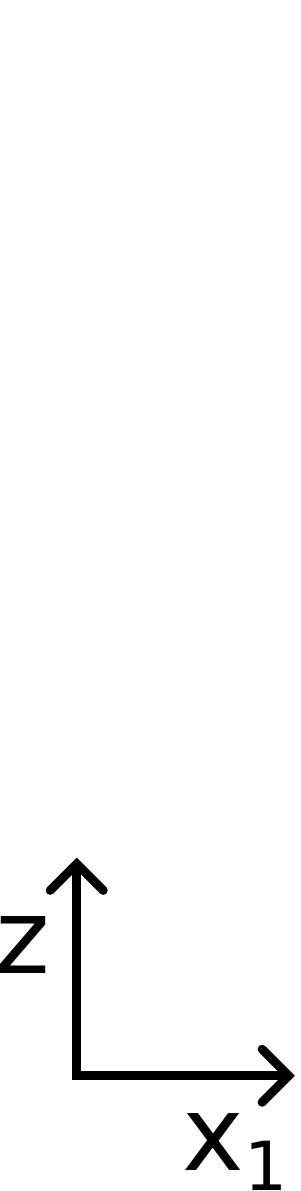}
	\includegraphics[height=120px]{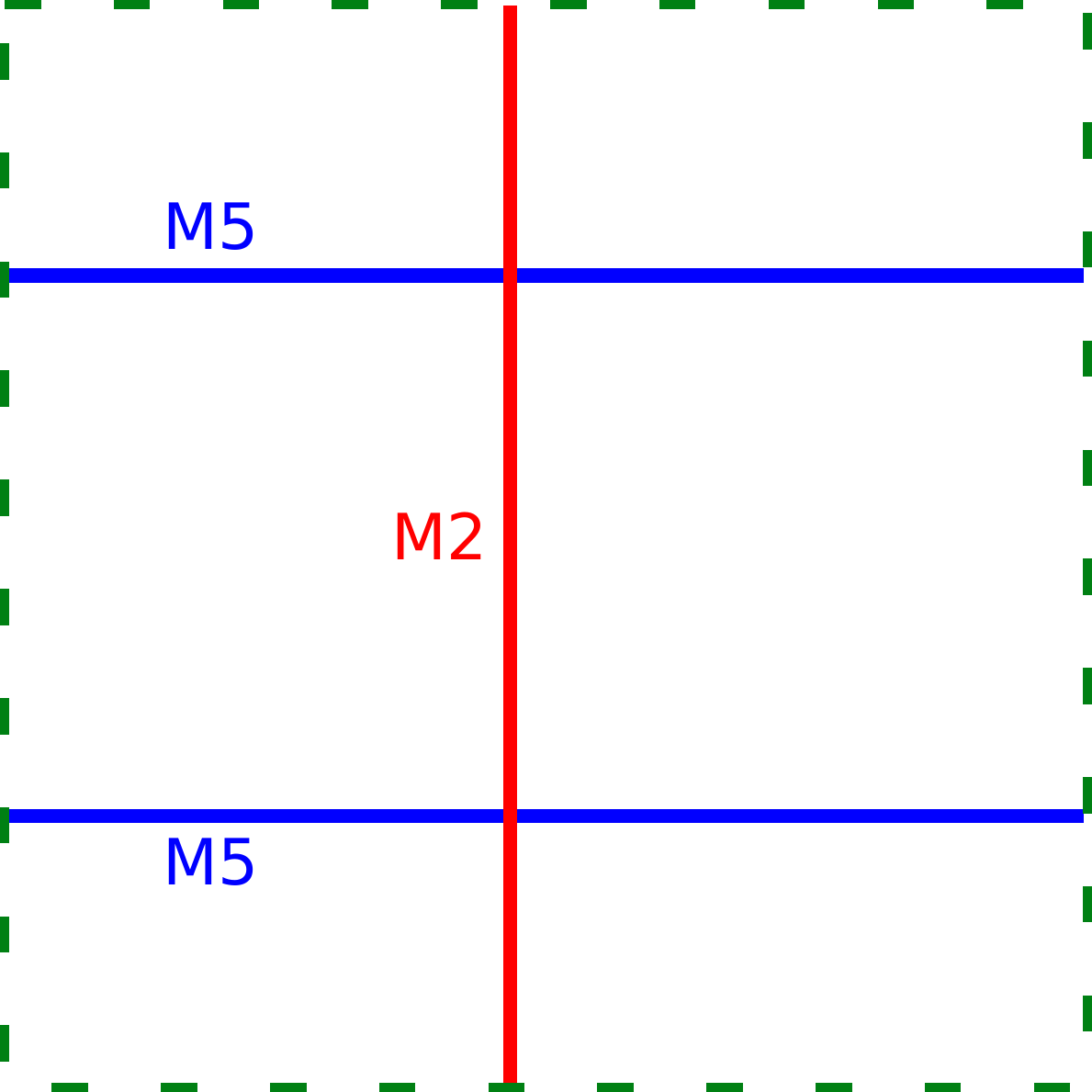}
	\hspace{.8em}
	\includegraphics[height=120px]{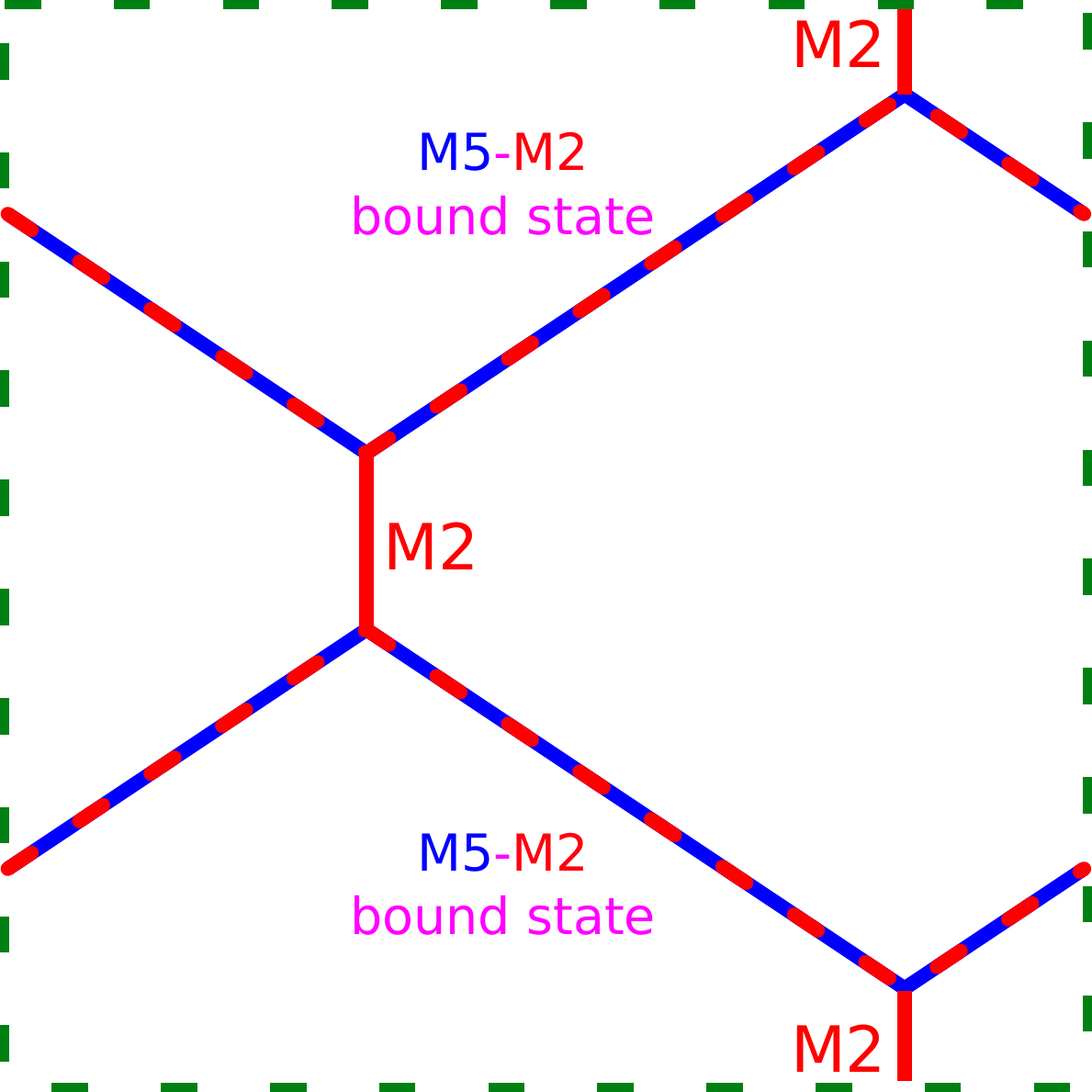}
	\hspace{.8em}
	\includegraphics[height=120px]{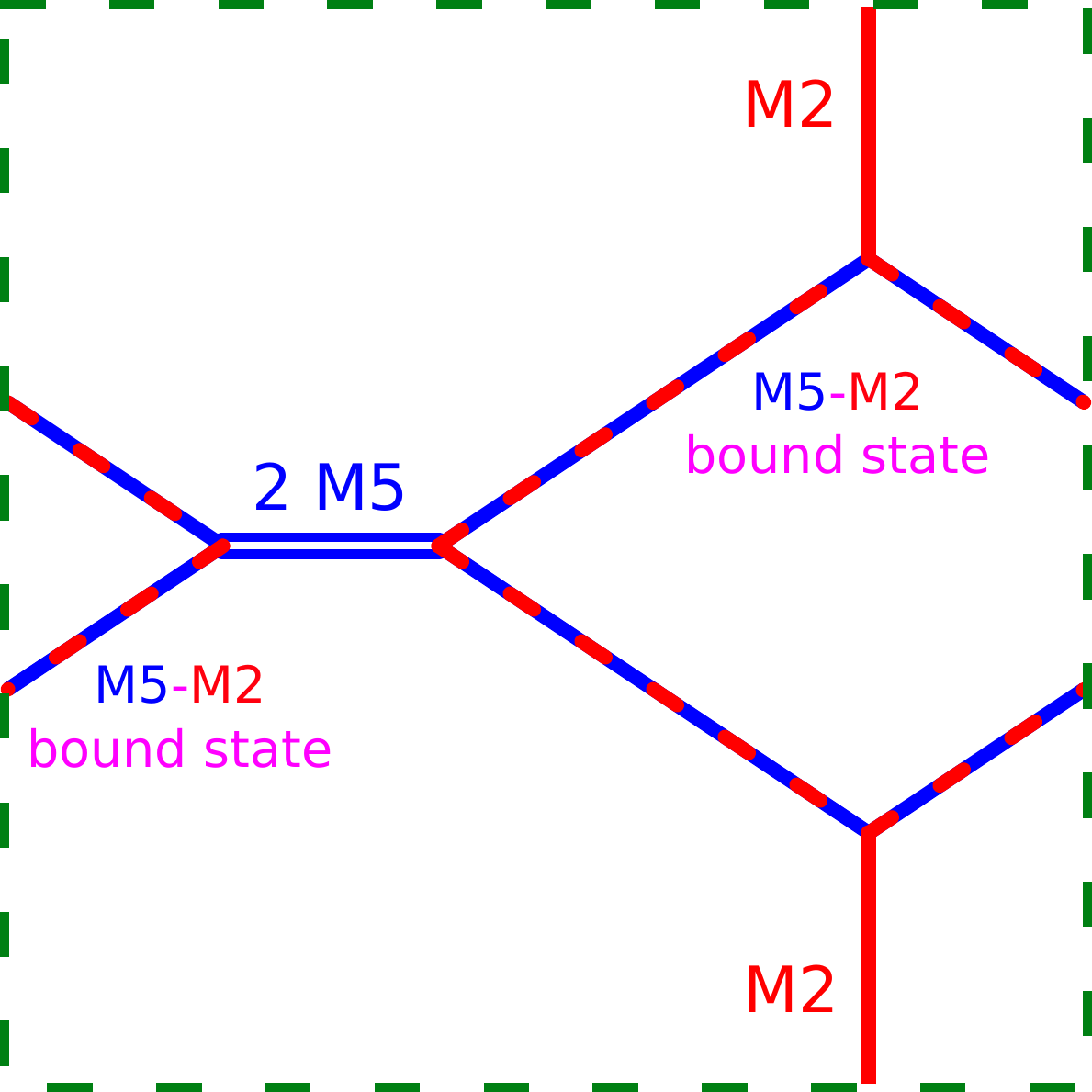}
	\caption{A super-maze made of 2 M5 branes and a single M2 brane which is smeared along three of the M5 brane worldvolume directions. Before the fractionation the M2 brane does not pull on the M5 branes, and can be freely taken away. After the fractionation (middle panel), each strip of the M2 branes deforms the M5 brane in its vicinity. As the branes move, the web depicted in the middle panel can also transform in the web depicted in the right panel, which has regions of coincident un-fluxed M5 branes. }
	\label{fig:flux_before_after}
\end{figure}

In a simplified model in which the M5-M2 furrow is smeared along 3 directions of the $T^4$, the supermaze becomes a string web made of M5 branes with M2 flux on it \cite{Bena:2022wpl}, see Fig. \ref{fig:flux_before_after}. These M5 branes wrap non-trivial cycles, so the backreaction of these simplified supermaze, after geometric transition, will involve topologically non-trivial bubbles in the compact dimensions.
Therefore, at the supergravity regime, while the mechanism that stabilises the superstrata's throat is the quantization of the angular momentum in the non-compact dimensions (see Section \ref{ssec:supertube_transition}), for the supermaze, one expects that the microstates do not degenerate into the black-hole solution thanks to the non-trivial topological bubbles all along the $x^{11}$ and $T^4$ directions.

\section{Questions, answers, and future directions}
\label{sec:questions}

\subsection{How does the super-maze evade horizon formation through the attractor mechanism?}

In supergravity, the fact that the asymptotic charges and global supersymmetries impose the near-horizon geometry is implemented by the \textit{attractor mechanism} (see e.g. \cite{Ferrara:1995ih}). However, the attractor mechanism follows an Ansatz which assumes two important hypotheses:
\begin{itemize}
    \item[1.] The solution is spherically symmetric in the external (\textit{i.e.} non-compact) spatial dimensions.%
    \footnote{Actually, the horizon does not need to have the geometry of a sphere; there can be attractor mechanisms for horizons with the geometry of a Riemann surface \cite{Bobev:2020jlb}. In such situations, the metric takes an Ansatz that preserves the symmetries of the Riemann surface as one moves away from the horizon. This is what really matters in the remaining of the discussion.}
    \item[2.] The solution is a warped product between the external and internal manifolds, but the warping depends only on the coordinates of the external dimensions, and not on the internal ones.
\end{itemize}

For most of microstate geometries families constructed in the literature -- for instance multi-centered/bubbling solutions and superstrata -- , the solutions evade the attractor mechanism by breaking the hypothesis 1, that is to say the spherical symmetry. 

If the supermaze admit a supergravity description, the way the supergravity solution would evade the attractor mechanism is probably by breaking the second hypothesis of the Ansatz. Indeed, before full reaction, locally at the scale of a M2-M5 furrow, the metric depends on the internal coordinates, in particular on the M-theory direction, $z$; see the solution in \cite{Lunin:2007mj}.

\subsection{Could the supermaze describe more entropy than that of the black hole?}

\textit{It seems from Fig. \ref{fig:M5-M2-P_wiggle} that a piece of M2-M5-P furrow has more degrees of freedom than a single strip of M2 brane ending on a M5 brane. Could the supermaze describe more entropy than the DVVM microstates, and thus have more entropy than the black hole?}
\vspace{1em}

First, the description in which one specifies the location of the wiggling furrow in the $(y,z,1)$ space gives a classical profile upon which one should \textit{in fine} apply geometric quantisation. Therefore, any additional degree of freedom the supermaze seems to have with respect to the DVVM microstates may not be independent from the `pure DVVM' degrees of freedom.
Besides, there is the possibility that the fermionic degrees of freedom of the DVVM microstates could be described in terms of some bosonic degrees of freedom of the supermaze that do not correspond to the motion of the M2 strips in the $T^4$.

Second, the projector only reveals what happens at a very local scale. There are additional constraints at the scale of the entire supermaze and at the scale of the M2/M5 furrow/spike. 

At the scale of the supermaze, take, for instance, the mode \eqref{coupled_M2M5_mode}, depicted in Fig. \ref{fig:M5-M2-P_wiggle}. One has to imagine this mode with a spherical symmetry in the $T^4$ directions, so that the mode corresponds to a M2 spike that becomes bigger and smaller, but at the same time the M5 brane has to move up and down. 
One has to check that the motion of the M5 brane is consistent with the motion of all the M2 strips that end on it, which reduces the number of degrees of freedom by a factor of $N_5$.

At the scale of the furrow/spike, one may think there is a characteristic size corresponding to the size of the base of the M2-M5 spike/furrow. But actually this size is controlled by the radius of the circle, $S_y^1$, the same way the typical size of a Callan-Maldacena spike is controlled by the string coupling, $g_s$. Below we write their connection through dualities.

\vspace{1em}
Consider a D4-brane in the directions 1234, and a F1-string along the direction $y$, ending on the D4-brane orthogonally. This picture is valid when $g_s \ll 1$. As one increases $g_s$, this system backreacts and undergoes the Callan-Maldacena effect: namely, the string pulls the world-volume of the D4-brane, thus forming a spike.

The M5-M2 bound state is dual to the D4-F1 Callan-Maldacena spike, after 11-dimensional uplift along $z$, and a flip in the coordinates $(y,z)$, in the following fashion:
\begin{align}
	\label{eq:DualChain_spike-furrow}
	\begin{pmatrix}
		\text{D4$(x^1x^2x^3x^4)$}\\
		\text{F1$(y)$}\\
	\end{pmatrix}_{\rm IIA}\hspace*{-0.4cm}
	\xrightarrow[]{~\text{uplift on}~z~}~
	\begin{pmatrix}
		\text{M5$(z,x^1x^2x^3x^4)$}\\
		\text{M2$(z,y)$}\\
	\end{pmatrix}_{\rm M}\hspace*{-0.4cm}
	\xrightarrow{~(z,y)\text{-flip}~}~
	\begin{pmatrix}
		\text{M5$(y,x^1x^2x^3x^4)$}\\
		\text{M2$(y,z)$}\\
	\end{pmatrix}_{\rm M}\hspace*{-0.4cm} \,.
\end{align}

At fixed number of branes and strings, the parameter that controls the backreaction of the Callan-Maldacena spike is the string coupling, $g_s$. From the duality chain (\ref{eq:DualChain_spike-furrow}), for a M5-M2 bound state, the transition between orthogonal M5-M2 branes into the M5-M2 furrow happens as the radius, $R_y$, of their common direction, $y$, increases. The backreaction parameter of the M5-M2 furrow, $R_y$, is linked with that of the D4-F1 Callan-Maldacena spike, $g_s$ by
\begin{align}
	\label{eq:R_y_VS_g_s}
	R_y=g_s \sqrt{\alpha'} \,, \qquad g_s=(R_y/l_{11})^{3/2} \,,
\end{align}
where $l_{11}$ is the eleven-dimensional Planck length.

In Type IIA string theory, as $g_s$ increases, the Callan-Maldacena spike grows in size in the directions of the D4, $(x^1,x^2,x^3,x^4)$. From equation (\ref{eq:R_y_VS_g_s}), as the radius $R_y$ increases (in units of the eleven-dimensional Planck length), the M5-M2 furrow gains more and more width -- that is to say, size in the $T^4$ directions.

For models of two-charge M5-M2 black holes, as well of three-charge M5-M2-P black holes, the radius of the common direction, $R_y$, is assumed to be important with respect to other compact directions, the torus $T^4$ and the M-theory circle $S_z$. This determines a large coupling constant, $g_s$, in Type IIA string theory, which implies an important pulling effect of the Callan-Maldacena spike. Therefore, one expects the M5-M2 furrow to be thick, and that the effect of the bound-state transition quite important for microstates of the M5-M2(-P) black hole.

\vspace{1em}
In a nutshell, there is a more global condition that fixes the size of the furrow at its base -- the size is controlled by $g_s$. Therefore, we think it is incorrect to interpret that the supermaze has degrees of freedom in addition of those of the DVVM microstates (at least not at the leading order in the entropy). Instead, the most natural interpretation should be that the DVVM microstates give a picture of the microstates at $g_s=0$, and so they are an \textit{approximation} of the supermaze, the same way the open string ending orthogonally on a D3 brane is the $g_s=0$ approximation of the Callan-Maldacena spike.

\subsection{Should local supersymmetry enhancement also work for D1-D5-P black holes?}

\textit{In your talk you enhanced the local supersymmetries of the DVVM microstates of the F1-NS5-P black hole. Should the LSE (local supersymmetry enhancement) principle also work for the microstates of the D1-D5-P black holes, often pictured as some 1-5 open strings?}
\vspace{1em}


Indeed, although the F1-NS5-P black hole is U-dual to the D1-D5-P black hole, the picture that we have of the microstates that account for their entropy are very different and not dual. A very interesting question is to understand whether one can find microstates with 16 local supersymmetries in all duality frames. 

One possible outcome is that LSE works for DVVM-type fractionation (where the entropy is coming from the fractionation of the F1's inside the NS5's), but not for `Strominger-Vafa'-type fractionation (where the entropy is coming from momentum-carrying 1-5 open strings). In other words, in this situation, one would be able to track the degrees of freedom characterising the microstates from $g_s=0$ (zero-gravity regime) to $g_s\lesssim 1/N$ (finite/weak-gravity regime) in the case of the F1-NS5-P black hole (with the supermaze degrees of freedom), but not in the case of the D1-D5-P black hole. If the goal was to prove that the horizon is an artefact of smearing the stringy degrees of freedom, one could be satisfied by this outcome: Since a horizon is a horizon in \textit{all} duality frames, one just needs to resolve the horizon in one of the duality frames.

Another outcome would be that one should be able to find degrees of freedom characterising the microstates that can be tracked from the zero-gravity regime to (at least) the finte/weak-gravity regime for the the D1-D5-P system as well.
These degrees of freedom do not need to be the 1-5 open strings, which perhaps cannot be tracked to the weak-gravity regime; but the hard part is to find enough of them to account for the Strominger-Vafa entropy \cite{Strominger:1996sh}. This is work in progress.


\vspace{1em}
Another type of supersymmetric black hole for which the local supersymmetric enhancement is unclear is those which involve branes of the same type orthogonal to each other.

The supersymmetry enhancement formalism just tells us what ingredients can be used in order to make a local 16-supersymmetric state. But the formalism does not tell about the \textit{size} of the bound state. For instance, take M2-M2 black holes, made of pairs of orthogonal M2 branes in e.g. a $T^4$, with one species of M2 branes wrapping the directions 12, and the other wrapping the directions 34. The glues to add in order to enhance the local supersymmetries are for example M2(13) and M2(24), or M2(14) and M2(23). The coefficients in front of the main excitations and the glues reveal that the interpolation from the M2(12) to the M2(34) has to be done through a holomorphic curve in the $T^4$; see \cite{Lunin:2008tf}. It is not clear that one has found a local resolution of the M2-M2 intersection. Instead, the LSE may well mean that, starting from the first M2(12), one needs to go around the scale of the $T^4$ to reach the second M2(34).

Similarly, a way to enhance the local supersymmetries of M5(y1234) and M5(y1256) can also be done through a holomorphic curve along the 3456 space. The LSE formalism does not tell if the size of the bound state is at the scale of the Calabi-Yau three-fold, like the MSW microstates \cite{Maldacena:1997de}, or if one can replace locally the orthogonal intersection between the M5(y1234) and M5(y1256) with some other configuration. 

\acknowledgments

I would like thank Iosif Bena, Davide Bufalini, Nejc \v{C}eplak, Alessandra Gnecchi, Dieter L\"ust, Valentin Reys, Nick Warner and all my collaborators in \cite{Bena:2022wpl} for interesting discussions.
This work is supported by by the German Research Foundation through a German-Israeli Project Cooperation (DIP) grant ``Holography and the Swampland.''

\appendix

\section{Projectors and involutions for branes}
\label{sec:projectors_and_involutions_for_branes}

In this Appendix we list the involutions associated to common brane type. In Type II string theory, they are:%
\begin{align}
	P_{\mathrm{ P  }} = \Gamma^{01}\,,&\qquad
	P_{\mathrm{ F1 }} = \Gamma^{01}\sigma_3 \,, \qquad
	\notag\\
	P_{\mathrm{ NS5}}^{\mathrm{ IIA}} = \Gamma^{012345} \,,& \qquad
	P_{\mathrm{ NS5}}^{\mathrm{ IIB}} = \Gamma^{012345} \sigma_3 \,,
	\notag\\		
	P_{\mathrm{ KKM(12345;6)}}^{\mathrm{ IIA}} = \Gamma^{012345} \sigma_3 \,, &\qquad
	P_{\mathrm{ KKM(12345;6)}}^{\mathrm{ IIB}} = \Gamma^{012345} \,,
	\\
	P_{\mathrm{ D0 }} = \Gamma^0 i\sigma_2 \,, \qquad
	P_{\mathrm{ D2 }} = \Gamma^{012} \sigma_1 \,,&\qquad 
	P_{\mathrm{ D4 }} = \Gamma^{01234} i\sigma_2 \,, \qquad
	P_{\mathrm{ D6 }} = \Gamma^{0123456} \sigma_1 \,,
	\notag\\
	P_{\mathrm{ D1 }} = \Gamma^{01} \sigma_1 \,, \qquad
	P_{\mathrm{ D3 }} = \Gamma^{0123} i\sigma_2 \,, &\qquad
	P_{\mathrm{ D5 }} = \Gamma^{012345} \sigma_1 \,.
	\notag
\end{align}

The projectors in M-theory are given by:
\begin{equation}
	P_{\mathrm{ P  }} = \Gamma^{01}\,, \qquad
	P_{\mathrm{ M2 }} = \Gamma^{012} \,, \qquad
	P_{\mathrm{ M5}} = \Gamma^{012345} \,, \qquad
	P_{\mathrm{ KKM(123456;7)}} = \Gamma^{0123456} \,.
\end{equation}

\bibliographystyle{JHEP}
\bibliography{proceeding}

\providecommand{\href}[2]{#2}\begingroup\raggedright\begin{thebibliography}{10}

\bibitem{Bena:2022wpl}
I.~Bena, S.D.~Hampton, A.~Houppe, Y.~Li and D.~Toulikas, \emph{{The (amazing)
  super-maze}}, \href{https://doi.org/10.1007/JHEP03(2023)237}{\emph{JHEP}
  {\bfseries 03} (2023) 237}
  [\href{https://arxiv.org/abs/2211.14326}{{\ttfamily 2211.14326}}].

\bibitem{Peet:2000hn}
A.W.~Peet, \emph{{TASI lectures on black holes in string theory}},
  \href{https://arxiv.org/abs/hep-th/0008241}{{\ttfamily hep-th/0008241}}.

\bibitem{Mathur:2005zp}
S.D.~Mathur, \emph{{The fuzzball proposal for black holes: An elementary
  review}}, \href{https://doi.org/10.1002/prop.200410203}{\emph{Fortsch. Phys.}
  {\bfseries 53} (2005) 793}
  [\href{https://arxiv.org/abs/hep-th/0502050}{{\ttfamily hep-th/0502050}}].

\bibitem{Lunin:2001fv}
O.~Lunin and S.D.~Mathur, \emph{{Metric of the multiply wound rotating
  string}}, \href{https://doi.org/10.1016/S0550-3213(01)00321-2}{\emph{Nucl.
  Phys.} {\bfseries B610} (2001) 49}
  [\href{https://arxiv.org/abs/hep-th/0105136}{{\ttfamily hep-th/0105136}}].

\bibitem{Lunin:2002iz}
O.~Lunin, J.M.~Maldacena and L.~Maoz, \emph{{Gravity solutions for the D1-D5
  system with angular momentum}},
  \href{https://arxiv.org/abs/hep-th/0212210}{{\ttfamily hep-th/0212210}}.

\bibitem{Kanitscheider:2006zf}
I.~Kanitscheider, K.~Skenderis and M.~Taylor, \emph{{Holographic anatomy of
  fuzzballs}}, \href{https://doi.org/10.1088/1126-6708/2007/04/023}{\emph{JHEP}
  {\bfseries 04} (2007) 023}
  [\href{https://arxiv.org/abs/hep-th/0611171}{{\ttfamily hep-th/0611171}}].

\bibitem{Taylor:2005db}
M.~Taylor, \emph{{General 2 charge geometries}}, {\emph{JHEP} {\bfseries 03}
  (2006) 009} [\href{https://arxiv.org/abs/hep-th/0507223}{{\ttfamily
  hep-th/0507223}}].

\bibitem{Rychkov:2005ji}
V.S.~Rychkov, \emph{{D1-D5 black hole microstate counting from supergravity}},
  \href{https://doi.org/10.1088/1126-6708/2006/01/063}{\emph{JHEP} {\bfseries
  01} (2006) 063} [\href{https://arxiv.org/abs/hep-th/0512053}{{\ttfamily
  hep-th/0512053}}].

\bibitem{Dabholkar:2004dq}
A.~Dabholkar, R.~Kallosh and A.~Maloney, \emph{{A stringy cloak for a classical
  singularity}},
  \href{https://doi.org/10.1088/1126-6708/2004/12/059}{\emph{JHEP} {\bfseries
  12} (2004) 059} [\href{https://arxiv.org/abs/hep-th/0410076}{{\ttfamily
  hep-th/0410076}}].

\bibitem{Sen:2004dp}
A.~Sen, \emph{{How does a fundamental string stretch its horizon?}},
  \href{https://doi.org/10.1088/1126-6708/2005/05/059}{\emph{JHEP} {\bfseries
  05} (2005) 059} [\href{https://arxiv.org/abs/hep-th/0411255}{{\ttfamily
  hep-th/0411255}}].

\bibitem{Hubeny:2004ji}
V.~Hubeny, A.~Maloney and M.~Rangamani, \emph{{String-corrected black holes}},
  \href{https://doi.org/10.1088/1126-6708/2005/05/035}{\emph{JHEP} {\bfseries
  05} (2005) 035} [\href{https://arxiv.org/abs/hep-th/0411272}{{\ttfamily
  hep-th/0411272}}].

\bibitem{Kanitscheider:2007wq}
I.~Kanitscheider, K.~Skenderis and M.~Taylor, \emph{{Fuzzballs with internal
  excitations}}, {\emph{JHEP} {\bfseries 06} (2007) 056}
  [\href{https://arxiv.org/abs/0704.0690}{{\ttfamily 0704.0690}}].

\bibitem{Bena:2011uw}
I.~Bena, J.~de~Boer, M.~Shigemori and N.P.~Warner, \emph{{Double, Double
  Supertube Bubble}},
  \href{https://doi.org/10.1007/JHEP10(2011)116}{\emph{JHEP} {\bfseries 10}
  (2011) 116} [\href{https://arxiv.org/abs/1107.2650}{{\ttfamily 1107.2650}}].

\bibitem{Mateos:2001qs}
D.~Mateos and P.K.~Townsend, \emph{{Supertubes}},
  \href{https://doi.org/10.1103/PhysRevLett.87.011602}{\emph{Phys. Rev. Lett.}
  {\bfseries 87} (2001) 011602}
  [\href{https://arxiv.org/abs/hep-th/0103030}{{\ttfamily hep-th/0103030}}].

\bibitem{Emparan:2001ux}
R.~Emparan, D.~Mateos and P.K.~Townsend, \emph{{Supergravity supertubes}},
  {\emph{JHEP} {\bfseries 07} (2001) 011}
  [\href{https://arxiv.org/abs/hep-th/0106012}{{\ttfamily hep-th/0106012}}].

\bibitem{Bena:2016ypk}
I.~Bena, S.~Giusto, E.J.~Martinec, R.~Russo, M.~Shigemori, D.~Turton et~al.,
  \emph{{Smooth horizonless geometries deep inside the black-hole regime}},
  \href{https://doi.org/10.1103/PhysRevLett.117.201601}{\emph{Phys. Rev. Lett.}
  {\bfseries 117} (2016) 201601}
  [\href{https://arxiv.org/abs/1607.03908}{{\ttfamily 1607.03908}}].

\bibitem{Bena:2017xbt}
I.~Bena, S.~Giusto, E.J.~Martinec, R.~Russo, M.~Shigemori, D.~Turton et~al.,
  \emph{{Asymptotically-flat supergravity solutions deep inside the black-hole
  regime}}, \href{https://doi.org/10.1007/JHEP02(2018)014}{\emph{JHEP}
  {\bfseries 02} (2018) 014}
  [\href{https://arxiv.org/abs/1711.10474}{{\ttfamily 1711.10474}}].

\bibitem{Bena:2022rna}
I.~Bena, E.J.~Martinec, S.D.~Mathur and N.P.~Warner, \emph{{Fuzzballs and
  Microstate Geometries: Black-Hole Structure in String Theory}},
  \href{https://arxiv.org/abs/2204.13113}{{\ttfamily 2204.13113}}.

\bibitem{Shigemori:2019orj}
M.~Shigemori, \emph{{Counting Superstrata}},
  \href{https://doi.org/10.1007/JHEP10(2019)017}{\emph{JHEP} {\bfseries 10}
  (2019) 017} [\href{https://arxiv.org/abs/1907.03878}{{\ttfamily
  1907.03878}}].

\bibitem{Mayerson:2020acj}
D.R.~Mayerson and M.~Shigemori, \emph{{Counting D1-D5-P microstates in
  supergravity}},
  \href{https://doi.org/10.21468/SciPostPhys.10.1.018}{\emph{SciPost Phys.}
  {\bfseries 10} (2021) 018}
  [\href{https://arxiv.org/abs/2010.04172}{{\ttfamily 2010.04172}}].

\bibitem{Bena:2014qxa}
I.~Bena, M.~Shigemori and N.P.~Warner, \emph{{Black-Hole Entropy from
  Supergravity Superstrata States}},
  \href{https://doi.org/10.1007/JHEP10(2014)140}{\emph{JHEP} {\bfseries 1410}
  (2014) 140} [\href{https://arxiv.org/abs/1406.4506}{{\ttfamily 1406.4506}}].

\bibitem{Li:2022apf}
Y.~Li, \emph{{The physics of black-hole microstates : what is the fate of the
  horizon?}}, Ph.D. thesis, Saclay, 2022.

\bibitem{Lin:2022rzw}
H.W.~Lin, J.~Maldacena, L.~Rozenberg and J.~Shan, \emph{{Holography for people
  with no time}},  \href{https://arxiv.org/abs/2207.00407}{{\ttfamily
  2207.00407}}.

\bibitem{Bena:2006kb}
I.~Bena, C.-W.~Wang and N.P.~Warner, \emph{{Mergers and Typical Black Hole
  Microstates}},
  \href{https://doi.org/10.1088/1126-6708/2006/11/042}{\emph{JHEP} {\bfseries
  11} (2006) 042} [\href{https://arxiv.org/abs/hep-th/0608217}{{\ttfamily
  hep-th/0608217}}].

\bibitem{Li:2021gbg}
Y.~Li, \emph{{Black holes and the swampland: the deep throat revelations}},
  \href{https://doi.org/10.1007/JHEP06(2021)065}{\emph{JHEP} {\bfseries 06}
  (2021) 065} [\href{https://arxiv.org/abs/2102.04480}{{\ttfamily
  2102.04480}}].

\bibitem{Li:2021utg}
Y.~Li, \emph{{An Alliance in the Tripartite Conflict over Moduli Space}},
  \href{https://arxiv.org/abs/2112.03281}{{\ttfamily 2112.03281}}.

\bibitem{deBoer:2008zn}
J.~de~Boer, S.~El-Showk, I.~Messamah and D.~Van~den Bleeken, \emph{{Quantizing
  N=2 Multicenter Solutions}},
  \href{https://doi.org/10.1088/1126-6708/2009/05/002}{\emph{JHEP} {\bfseries
  05} (2009) 002} [\href{https://arxiv.org/abs/0807.4556}{{\ttfamily
  0807.4556}}].

\bibitem{Bena:2022sge}
I.~Bena, N.~Ceplak, S.~Hampton, Y.~Li, D.~Toulikas and N.P.~Warner,
  \emph{{Resolving Black-Hole Microstructure with New Momentum Carriers}},
  \href{https://arxiv.org/abs/2202.08844}{{\ttfamily 2202.08844}}.

\bibitem{Callan:1997kz}
C.G.~Callan and J.M.~Maldacena, \emph{{Brane death and dynamics from the
  Born-Infeld action}},
  \href{https://doi.org/10.1016/S0550-3213(97)00700-1}{\emph{Nucl. Phys. B}
  {\bfseries 513} (1998) 198}
  [\href{https://arxiv.org/abs/hep-th/9708147}{{\ttfamily hep-th/9708147}}].

\bibitem{Izquierdo:1995ms}
J.M.~Izquierdo, N.D.~Lambert, G.~Papadopoulos and P.K.~Townsend, \emph{{Dyonic
  membranes}}, \href{https://doi.org/10.1016/0550-3213(95)00606-0}{\emph{Nucl.
  Phys. B} {\bfseries 460} (1996) 560}
  [\href{https://arxiv.org/abs/hep-th/9508177}{{\ttfamily hep-th/9508177}}].

\bibitem{Lunin:2007mj}
O.~Lunin, \emph{{Strings ending on branes from supergravity}},
  \href{https://doi.org/10.1088/1126-6708/2007/09/093}{\emph{JHEP} {\bfseries
  09} (2007) 093} [\href{https://arxiv.org/abs/0706.3396}{{\ttfamily
  0706.3396}}].

\bibitem{Seiberg:1997zk}
N.~Seiberg, \emph{{New theories in six-dimensions and matrix description of M
  theory on T**5 and T**5 / Z(2)}},
  \href{https://doi.org/10.1016/S0370-2693(97)00805-8}{\emph{Phys. Lett. B}
  {\bfseries 408} (1997) 98}
  [\href{https://arxiv.org/abs/hep-th/9705221}{{\ttfamily hep-th/9705221}}].

\bibitem{Kutasov:2001uf}
D.~Kutasov, \emph{{Introduction to little string theory}}, {\emph{ICTP Lect.
  Notes Ser.} {\bfseries 7} (2002) 165}.

\bibitem{Dijkgraaf:1996cv}
R.~Dijkgraaf, E.P.~Verlinde and H.L.~Verlinde, \emph{{BPS spectrum of the
  five-brane and black hole entropy}},
  \href{https://doi.org/10.1016/S0550-3213(96)00638-4}{\emph{Nucl. Phys. B}
  {\bfseries 486} (1997) 77}
  [\href{https://arxiv.org/abs/hep-th/9603126}{{\ttfamily hep-th/9603126}}].

\bibitem{Maldacena:1996ya}
J.M.~Maldacena, \emph{{Statistical entropy of near extremal five-branes}},
  \href{https://doi.org/10.1016/0550-3213(96)00368-9}{\emph{Nucl. Phys. B}
  {\bfseries 477} (1996) 168}
  [\href{https://arxiv.org/abs/hep-th/9605016}{{\ttfamily hep-th/9605016}}].

\bibitem{Ferrara:1995ih}
S.~Ferrara, R.~Kallosh and A.~Strominger, \emph{{N=2 extremal black holes}},
  \href{https://doi.org/10.1103/PhysRevD.52.R5412}{\emph{Phys. Rev.} {\bfseries
  D52} (1995) R5412} [\href{https://arxiv.org/abs/hep-th/9508072}{{\ttfamily
  hep-th/9508072}}].

\bibitem{Bobev:2020jlb}
N.~Bobev, F.F.~Gautason and K.~Parmentier, \emph{{Holographic Uniformization
  and Black Hole Attractors}},
  \href{https://doi.org/10.1007/JHEP06(2020)095}{\emph{JHEP} {\bfseries 06}
  (2020) 095} [\href{https://arxiv.org/abs/2004.05110}{{\ttfamily
  2004.05110}}].

\bibitem{Strominger:1996sh}
A.~Strominger and C.~Vafa, \emph{{Microscopic Origin of the Bekenstein-Hawking
  Entropy}}, \href{https://doi.org/10.1016/0370-2693(96)00345-0}{\emph{Phys.
  Lett.} {\bfseries B379} (1996) 99}
  [\href{https://arxiv.org/abs/hep-th/9601029}{{\ttfamily hep-th/9601029}}].

\bibitem{Lunin:2008tf}
O.~Lunin, \emph{{Brane webs and 1/4-BPS geometries}},
  \href{https://doi.org/10.1088/1126-6708/2008/09/028}{\emph{JHEP} {\bfseries
  09} (2008) 028} [\href{https://arxiv.org/abs/0802.0735}{{\ttfamily
  0802.0735}}].

\bibitem{Maldacena:1997de}
J.M.~Maldacena, A.~Strominger and E.~Witten, \emph{{Black hole entropy in
  M-theory}}, {\emph{JHEP} {\bfseries 12} (1997) 002}
  [\href{https://arxiv.org/abs/hep-th/9711053}{{\ttfamily hep-th/9711053}}].

\end{thebibliography}\endgroup


\end{document}